\documentclass[letterpaper,11pt]{article}

\usepackage[utf8]{inputenc}
\usepackage[T1]{fontenc}
\usepackage{cfr-lm}

\usepackage{stmaryrd}
\usepackage{cite}
\usepackage[dvips]{graphicx}
\usepackage{graphpap}
\usepackage{ifthen}
\usepackage{enumerate}
\usepackage{amssymb}
\usepackage{mathrsfs}
\usepackage[errorshow]{tracefnt}
\usepackage{fancybox}
\usepackage{amsfonts}
\usepackage{amsmath}
\usepackage{amssymb}
\usepackage{multirow}
\usepackage{array}
\usepackage{mathtools}
\usepackage{soul}
\usepackage{pict2e}
\usepackage{mathabx}
\DeclareMathAlphabet{\mathpzc}{OT1}{pzc}{m}{it}

\usepackage{lscape}
\usepackage{xypic}
\usepackage{color}
\usepackage[
color,matrix,arrow]{xy}
\usepackage{setspace}

\usepackage{enumitem}

\usepackage{graphicx}

\makeatletter
\newcommand{\thickhline}{%
    \noalign {\ifnum 0=`}\fi \hrule height 1pt
    \futurelet \reserved@a \@xhline
}
\newcolumntype{'}{@{\hskip\tabcolsep\vrule width 1pt\hskip\tabcolsep}}
\makeatother
\newcolumntype{"}{@{\hskip\tabcolsep\vrule width 1.5pt\hskip\tabcolsep}}
\makeatother

\newcommand{\scr}{\mathscr}

\def\ie{{\it i.e.}}

\def\eg{{\it e.g.}}

\def\boxit#1{\vbox{\hrule\hbox{\vrule\kern3pt
             \vbox{\kern3pt#1\kern3pt}\kern3pt\vrule}\hrule}}

\newcommand{\Blue}[1]{{\color{blue} #1}}

\newcommand{\beq}{\begin{equation}}
\newcommand{\beqn}{\begin{equation*}}
\newcommand{\eeq}{\end{equation}}
\newcommand{\eeqn}{\end{equation*}}
\newcommand{\beqa}{\begin{eqnarray}}
\newcommand{\beqan}{\begin{eqnarray*}}
\newcommand{\eeqa}{\end{eqnarray}}
\newcommand{\eeqan}{\end{eqnarray*}}
\newcommand{\bdm}{\begin{displaymath}}
\newcommand{\edm}{\end{displaymath}}

\newcommand{\ba}{\begin{array}}
\newcommand{\ea}{\end{array}}

\newcommand\nn{\nonumber}

\newcommand\benu{\begin{enumerate}}
\newcommand\eenu{\end{enumerate}}
\newcommand\bit{\begin{itemize}}
\newcommand\eit{\end{itemize}}

\def\dim{\mathrm{dim\,}}
\def\sdim{\mathrm{sdim\,}}

\def\der'{\mathfrak{der}'\,}
\def\der{\mathfrak{der}\,}
\def\str'{\mathfrak{str}'\,}
\def\str{\mathfrak{str}\,}

\def\gl{\mathfrak{gl}}

\newcommand{\de}{\delta}

\newcommand{\ediagram}{
\begin{picture}(330,70)(45,-10)
\put(48,-10){${\gamma_{-1}}$}
\put(88,-10){${\gamma_{0}}$}
\put(128,-10){${\gamma_{1}}$}
\put(242,-10){${\gamma_{r-4}}$}
\put(282,-10){${\gamma_{r-3}}$}
\put(322,-10){${\gamma_{r-2}}$}
\put(362,-10){${\gamma_{r-1}}$}
\put(303,48){${\gamma_{r}}$}
\thicklines
\put(50,10){\line(1,1){3.5}}
\put(50,10){\line(-1,1){3.5}}
\put(50,10){\line(1,-1){3.5}}
\put(50,10){\line(-1,-1){3.5}}
\multiput(50,10)(40,0){3}{\circle{10}}
\multiput(250,10)(40,0){4}{\circle{10}}
\multiput(55,10)(40,0){2}{\line(1,0){30}}
\put(135,10){\line(1,0){20}}
\put(165,10){\line(1,0){10}}
\put(185,10){\line(1,0){10}}
\put(205,10){\line(1,0){10}}
\put(225,10){\line(1,0){20}}
\put(255,10){\line(1,0){30}}
\put(295,10){\line(1,0){30}}
\put(335,10){\line(1,0){30}}
\put(290,50){\circle{10}}
\put(290,15){\line(0,1){30}}
\end{picture}
}

\newcommand{\ediagramTwoGreyNodes}{
\begin{picture}(330,70)(45,-10)
\put(48,-10){${\beta_{-1}}$}
\put(88,-10){${\beta_{0}}$}
\put(128,-10){${\beta_{1}}$}
\put(242,-10){${\beta_{r-4}}$}
\put(282,-10){${\beta_{r-3}}$}
\put(322,-10){${\beta_{r-2}}$}
\put(362,-10){${\beta_{r-1}}$}
\put(303,48){${\beta_{r}}$}
\thicklines
\put(50,10){\line(1,1){3.5}}
\put(50,10){\line(-1,1){3.5}}
\put(50,10){\line(1,-1){3.5}}
\put(50,10){\line(-1,-1){3.5}}
\put(90,10){\line(1,1){3.5}}
\put(90,10){\line(-1,1){3.5}}
\put(90,10){\line(1,-1){3.5}}
\put(90,10){\line(-1,-1){3.5}}
\multiput(50,10)(40,0){3}{\circle{10}}
\multiput(250,10)(40,0){4}{\circle{10}}
\multiput(55,10)(40,0){2}{\line(1,0){30}}
\put(135,10){\line(1,0){20}}
\put(165,10){\line(1,0){10}}
\put(185,10){\line(1,0){10}}
\put(205,10){\line(1,0){10}}
\put(225,10){\line(1,0){20}}
\put(255,10){\line(1,0){30}}
\put(295,10){\line(1,0){30}}
\put(335,10){\line(1,0){30}}
\put(290,50){\circle{10}}
\put(290,15){\line(0,1){30}}
\end{picture}
}

\newcommand{\tf}{\tfrac}

\newcommand{\dlb}{\ensuremath{[\![}}
\newcommand{\drb}{\ensuremath{]\!]}}

\newcommand{\ad}{\text{ad}\,}

\def\fg{{\mathfrak g}}

\def\sB{{\scr B}}

\def\sh{\sharp}
\def\fl{\flat}

\def\*{\partial}

\def\tE{\widetilde E}
\def\tF{\widetilde F}
\def\tk{\widetilde k}
\def\tR{\widetilde R}
\def\ttR{\widetilde{\widetilde R}}
\def\tf{\widetilde f}

\def\tU{\widetilde U}
\def\tV{\widetilde V}

\def\LL{{\scr L}}

\RequirePackage{hyperref}

\numberwithin{equation}{section}

\begin{document}

\frenchspacing

\null\vspace{-18mm}

\includegraphics[width=2cm]{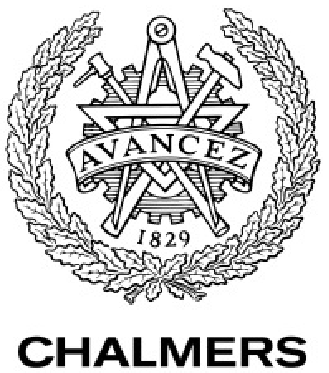}

\vspace{-17mm}
{\flushright Gothenburg preprint \\ April, 2018\\}

\vspace{5mm}

\hrule

\vspace{12mm}


\pagestyle{empty}

\begin{center}
  {\LARGE \bf \sc $L_{\infty}$ algebras for extended geometry}
\\[4mm]
  {\LARGE \bf \sc from Borcherds superalgebras}
    \\[12mm]

{\large
Martin Cederwall and Jakob Palmkvist}

\vspace{12mm}
       {\it Division for Theoretical Physics, Department of Physics,\\
         Chalmers University of Technology,
 SE-412 96 Gothenburg, Sweden}

\end{center}

\vspace{15mm}


\begin{quote}

\textbf{Abstract:}
We examine the structure of gauge transformations in extended
geometry, the framework unifying double geometry, exceptional
geometry, etc. This is done by giving the variations of the
ghosts in a
Batalin--Vilkovisky framework, or
equivalently, an $L_\infty$ algebra.
The $L_\infty$ brackets are given as derived brackets constructed
using an underlying Borcherds superalgebra ${\scr B}({\mathfrak g}_{r+1})$,
which is a double extension of the structure algebra ${\mathfrak g}_r$.
The construction includes a set of ``ancillary'' ghosts.
All
brackets involving the infinite sequence
of ghosts are given explicitly. All even brackets above the $2$-brackets
vanish, and the coefficients appearing in the brackets are given by
Bernoulli numbers.
The results are valid in the absence of ancillary transformations at
ghost number $1$.  We present evidence that in order to go further,
the underlying algebra should be the corresponding tensor hierarchy algebra.

\end{quote} 
\vfill

\hrule

\noindent{\tiny email:
  martin.cederwall@chalmers.se, jakob.palmkvist@chalmers.se}

\newpage

\tableofcontents

\pagestyle{plain}

\section{Introduction}

The ghosts in exceptional field theory
\cite{Berman:2012vc}, and generally in extended field theory
with an extended structure algebra $\fg_r$ \cite{Cederwall:2017fjm},
are known to fall into ${\scr B}_+(\fg_r)$, 
the positive levels of a Borcherds
superalgebra ${\scr B}(\fg_r)$
\cite{Palmkvist:2015dea,Cederwall:2015oua}.
We use the concept of ghosts, including ghosts for ghosts etc., as a
convenient tool to encode the structure of the gauge symmetry
(structure constants, reducibility and so on) in a
classical field theory using the (classical) Batalin--Vilkovisky framework.

It was shown in ref. \cite{Palmkvist:2015dea} how generalised
diffeomorphisms for $E_r$ have a natural formulation in terms of the
structure constants of the Borcherds superalgebra
${\scr B}(E_{r+1})$.
This generalises to extended geometry in general
\cite{Cederwall:2017fjm}.
The more precise r\^ole of the Borcherds superalgebra has not been
spelt out, and one of the purposes of the present paper is to fill this gap.
The gauge structure of extended geometry will be described as an $L_\infty$
algebra, governed by an underlying Borcherds
superalgebra ${\scr B}(\fg_{r+1})$. The superalgebra ${\scr
  B}(\fg_{r+1})$ generalises
${\scr B}(E_{r+1})$ in ref. \cite{Palmkvist:2015dea}, and
is obtained from the structure algebra $\fg_r$ by adding two more
nodes to the Dynkin diagram, 
as will be explained in Section \ref{BorcherdsSection}.
In cases where the superalgebra is finite-dimensional, such as double
field theory
\cite{Tseytlin:1990va,Siegel:1993xq,Siegel:1993bj,Hitchin:2010qz,Hull:2004in,Hull:2006va,Hull:2009mi,Hohm:2010jy,Hohm:2010pp,Jeon:2012hp,Park:2013mpa,Berman:2014jba,Cederwall:2014kxa,Cederwall:2014opa,Cederwall:2016ukd},
the structure simplifies to an $L_{n<\infty}$ algebra
\cite{Deser:2016qkw,Hohm:2017pnh,Deser:2018oyg},
and the reducibility becomes finite.

It is likely that a consistent treatment of quantum extended
  geometry will require a full Batalin--Vilkovisky treatment of the
  ghost sector, which is part of the motivation behind our
  work. Another, equally strong motivation is the belief that the
  underlying superalgebras carry much information about the models ---
  also concerning physical fields and their dynamics ---
  and that this can assist us in the future when investigating
  extended geometries bases on infinite-dimensional structure algebras.

The first $8-r$ levels in ${\scr B}(E_r)$
consist of $E_r$-modules for form fields
in exceptional field theory
\cite{Hull:2007zu,Pacheco:2008ps,Hillmann:2009pp,Berman:2010is,Berman:2011pe,Coimbra:2011ky,Coimbra:2012af,Berman:2012vc,Park:2013gaj,Cederwall:2013naa,Cederwall:2013oaa,Aldazabal:2013mya,Hohm:2013pua,Blair:2013gqa,Hohm:2013vpa,Hohm:2013uia,Hohm:2014fxa,Cederwall:2015ica,Bossard:2017aae},
locally describing eleven-dimensional supergravity.
Inside this window, there is a
connection-free but covariant derivative, taking an element in $R_p$
at level $p$ to $R_{p-1}$ at level $p-1$
\cite{Cederwall:2013naa}. Above the window, the modules, when
decomposed as $\mathfrak{gl}(r)$ modules with respect to a local choice
of section, start to contain mixed tensors, and covariance is lost.
For $E_8$, the window closes, not even the generalised
diffeomorphisms are covariant \cite{Cederwall:2015ica} and there are
additional restricted local $E_8$ transformations \cite{Hohm:2014fxa}.
Such transformations were named ``ancillary'' in ref.
\cite{Cederwall:2017fjm}. In the present paper, we will not treat the situation 
where ancillary transformations arise in the commutator of two
generalised diffeomorphisms, but we will extend the concept of
ancillary ghosts to higher ghost number. It will become clear from the
structure of the doubly extended Borcherds superalgebra
${\scr B}(\fg_{r+1})$ why and when such extra restricted ghosts
appear, and what their precise connection to \eg\ the loss of
covariance is.

A by-product of our construction is that all identities previously
derived on a case-by-case basis, relating to the ``form-like''
properties of the elements in the tensor hierarchies
\cite{Cederwall:2013naa,Wang:2015hca}, are derived in a
completely general manner.

Although the exceptional geometries are the most interesting cases
where the structure has not yet been formulated, we will perform all
our calculations in the general setting with arbitrary structure group
(which for simplicity will be taken to be simply laced, although
non-simply laced groups present no principal problem). The general
formulation of ref. \cite{Cederwall:2017fjm} introduces no additional
difficulty compared to any special case,
and in fact provides the best unifying formalism also for
the different exceptional groups.
We note that the gauge symmetries of exceptional generalised
geometry have been dealt
with in the $L_\infty$ algebra framework earlier
\cite{Baraglia:2011dg}. However, this was done in terms of
a formalism where ghosts are not 
collected into modules of $E_r$, but consist of the diffeomorphism
parameter together with
forms for the ghosts of the tensor gauge transformations (\ie, in
generalised geometry, not in extended geometry).

In Section \ref{BorcherdsSection}, details about the Borcherds superalgebra 
${\scr B}(\fg_{r+1})$ are given. Especially, the double grading
relevant for our purposes is introduced, and the (anti-)commutators
are given in this basis. Section \ref{SectionLieDetSection} introduces
the generalised Lie derivative and the section constraint in terms of
the Borcherds superalgebra bracket. In Section
\ref{OperatorsSection} we show how the generalised Lie derivative
arises naturally from a nilpotent derivative on the
${\scr B}(\fg_{r})$ subalgebra, and how ancillary terms/ghosts fit into the
algebraic structure. Some further operators related to ancillary terms
are introduced, and identities between the operators are derived.
Section \ref{BVLInftySection} is an interlude concerning $L_\infty$
algebras and Batalin--Vilkovisky ghosts.
The non-ancillary part of the $L_\infty$ brackets, \ie, the part where
ghosts and brackets belong 
to the ${\scr B}_+(\fg_{r})$ subalgebra, is derived in
Section \ref{NonAncillaryBracketsSection}.
The complete non-ancillary variation $(S,C)=\sum_{n=1}^\infty\dlb C^n\drb$ can
formally be written as
\begin{align}
(S,C)=dC+g(\ad C)\LL_CC\;,
\end{align}
where $g$ is the function
\begin{align}
g(x)=\frac2{1-e^{-2x}}-\frac1x\;,
\end{align}
containing Bernoulli numbers in its Maclaurin series.
Ancillary ghosts are
introduced in Section \ref{AncillaryGhostsSection}, and the complete
structure of the $L_\infty$ brackets is presented in Section
\ref{FullBVSection}. Some examples, including ordinary diffeomorphisms (the
algebra of vector fields), double diffeomorphisms and exceptional
diffeomorphisms, are given in Section \ref{ExamplesSection}. We
conclude with a discussion, with focus on the extension of the present
construction to situations where ancillary transformations are present
already in the commutator of two generalised diffeomorphisms.

\section{The Borcherds superalgebra\label{BorcherdsSection}}

For simplicity we assume the structure algebra $\fg_r$ to be simply laced,
and we normalise the inner product in the real root space by
$(\alpha_i,\alpha_i)=2$. We let the coordinate module, which we denote
$R_1=R(-\lambda)$, be a lowest
weight module\footnote{In
  refs. \cite{Bossard:2017aae,Cederwall:2017fjm}, the coordinate
  module was taken to be a highest weight module. We prefer to reverse
these conventions (in agreement with ref. \cite{Palmkvist:2015dea}).
With the standard basis of simple roots in the superalgebra, its
positive levels consists of {\it lowest} weight $\fg_r$-modules.
In the present paper the distinction is not essential, since the cases
treated all concern finite-dimensional $\fg_r$ and finite-dimensional $\fg_r$-modules.}
with lowest weight $-\lambda$. Then the derivative module is a highest
weight module $R(\lambda)$ with highest weight $\lambda$, and
$R(-\lambda)=\overline{R(\lambda)}$.

As explained in ref. \cite{Palmkvist:2015dea} we can extend $\fg_r$ to a Lie algebra $\fg_{r+1}$
or to a Lie superalgebra
$\scr B(\fg_r)$
by adding a node to the Dynkin diagram. In the first case, the
additional node is an ordinary ``white'' node,  
the corresponding simple root
$\alpha_0$ satisfies $(\alpha_0,\alpha_0)=2$, and the resulting Lie
algebra $\fg_{r+1}$ is a Kac--Moody algebra like $\fg_r$ itself. In the
second case, 
the additional node is ``grey'', corresponding to a simple root
$\beta_0$. It satisfies $(\beta_0,\beta_0)=0$, and is furthermore a fermionic (\ie, odd)
root, which means that the associated Chevalley generators $e_0$
and $f_0$ belong to 
the fermionic subspace of the resulting Lie superalgebra $\scr B(\fg_r)$. In
both cases, the inner product of the additional simple root with those
of $\fg_r$ is given by the Dynkin labels of $\lambda$, with a minus
sign, 
\begin{align}
-\lambda_i=-(\lambda,\alpha_i)=(\alpha_0,\alpha_i)=(\beta_0,\beta_i)\;,
\end{align}
where we have set $\alpha_i = \beta_i$ ($i=1,2,\ldots,r$).

\begin{figure}
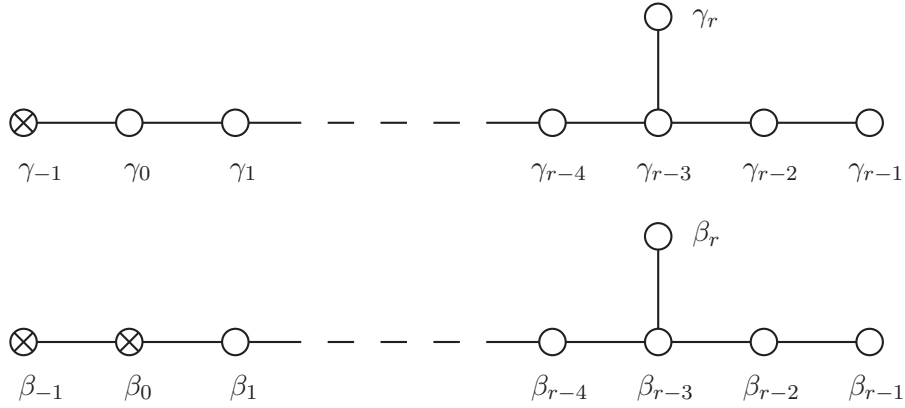

\begin{center}
\ediagram 
\end{center}
\begin{center}
\ediagramTwoGreyNodes
\end{center}
\caption{\label{DynkinFigure}\it Dynkin diagrams of $\scr B(E_{r+1})$
  together with our notation for the simple roots represented by the
  nodes.} 
\end{figure}

We can extend $\fg_{r+1}$ and $\scr B(\fg_r)$ further to
a Lie superalgebra $\scr B(\fg_{r+1})$ by adding one more node to the
Dynkin diagrams.\footnote{In ref. \cite{Cederwall:2017fjm}, the algebras $\fg_{r+1}$, $\scr B(\fg_r)$ and $\scr B(\fg_{r+1})$ were called
$\scr A$, $\scr B$ and $\scr C$, respectively.}
We will then get two different Dynkin diagrams (two different sets of
simple roots) corresponding to the same 
Lie superalgebra $\scr B(\fg_{r+1})$. These are shown in Figure
\ref{DynkinFigure} in the case when $\fg=E_r$ and $\lambda$ is the
highest weight of the 
derivative module in exceptional geometry.
The line between the two grey nodes in the second diagram indicate
that the inner product of the two corresponding simple roots is
$(\beta_{-1},\beta_0)=1$, 
not $-1$ as when one or both of the nodes are white.

The two sets of simple roots are related to each other by
\begin{align}
\gamma_{-1} &=-\beta_{-1}\;,&  
\gamma_{0} &=\beta_{-1}+\beta_0\;,  &   
\gamma_{i} &=\beta_{i}\;.
\end{align}
This corresponds to a ``generalised Weyl transformation'' or
``odd Weyl reflection''
\cite{Dobrev:1985qz}, which provides a  map between the two sets of Chevalley
generators mapping the 
defining relations to each other, thus inducing an isomorphism.

In spite of the notation $\scr B(\fg_{r+1})$ we choose to consider this algebra as constructed
from the second Dynkin diagram in Figure \ref{DynkinFigure}, which means that we let $e_0$, $f_0$ and $h_0$
be associated to $\beta_0$ rather than $\gamma_0$. For $\beta_{-1}$, we drop the subscript and write the associated generators
simply as $e$, $f$ and $h$.
They satisfy the (anti-)commutation relations
\begin{align}
[h,e]=[h,f]&=0\;&[e,f]&=h\;. \label{superheisenberg}
\end{align}
Acting with $h$ on $e_0$ and $f_0$ we have
\begin{align}
[h,e_0]&=e_0\;,& [h,f_0]&=-f_0\;.
\end{align}
Throughout the paper the notation $[\cdot,\cdot]$ is used for the
Lie super-bracket of the superalgebra, disregarding the statistics of
the generators. Thus, we do not use a separate notation
(\eg\ $\{\cdot,\cdot\}$, common in the physics literature) for
brackets between a pair of fermionic elements.

Let $k$ be an element in the Cartan subalgebra of $\scr B(\fg_r)$ that commutes with $\fg_r$ and satisfies $[k,e]=e$ and $[k,f]=-f$
when we extend $\scr B(\fg_r)$ to $\scr B(\fg_{r+1})$. In the Cartan subalgebra of $\scr B(\fg_{r+1})$, set $\tk = k+h$, so that $[e,f]=h=\tk-k$.
We then have
\begin{align}
[k,e_0] &= -(\lambda,\lambda)e_0\;, & [k,e] &= e\;,\nn\\
[k,f_0] &= (\lambda,\lambda)f_0\;, & [k,f] &= -f\;,
\end{align}
\begin{align}
[\tk,e_0] &= (1-(\lambda,\lambda))e_0\;, & [\tk,e] &= e\;,\nn\\
[\tk,f_0] &= ((\lambda,\lambda)-1)f_0\;, & [\tk,f] &= -f\;.
\end{align}

The Lie superalgebra $\scr B(\fg_{r+1})$
can be given a $(\mathbb{Z} \times \mathbb{Z})$-grading
with respect to $\beta_{0}$ and $\beta_{-1}$. It is then decomposed
into a direct sum of $\fg_r$ modules 
\begin{align}
\scr B(\fg_{r+1}) = \bigoplus_{(p,q) \in \mathbb{Z} \times \mathbb{Z} } R_{(p,q)}\;,
\end{align}
where $R_{(p,q)}$ is spanned by root vectors (together with the Cartan
generators if $p=q=0$) such that the corresponding roots have coefficients
$p$ and $q$ for $\beta_0$ and $\beta_{-1}$, respectively, when expressed as linear combinations of the simple roots.
We will refer 
to the degrees $p$ and $q$ as {\it level} and {\it height},
respectively. They are the eigenvalues of the adjoint action of $h=\tk
- k$ and 
the Cartan element 
\begin{align} \label{q-eigenvalue}
q=(1-(\lambda,\lambda))k+(\lambda,\lambda)\tk=k+(\lambda,\lambda)h\;,
\end{align}
respectively. Thus
\begin{align}
[q,e_0] &= [q,f_0] =0\;, & [q,e] &= e\;,&[q,f] &= -f\;.
\end{align}
In the same way as the Lie superalgebra $\scr B(\fg_{r+1})$ can be
decomposed with 
respect to $\beta_{0}$ and $\beta_{-1}$, it can also be decomposed with
respect to $\gamma_{0}$ and $\gamma_{-1}$.
Then the degrees $m$ and $n$, corresponding to $\gamma_0$ and
$\gamma_{-1}$, respectively, are related to the level and height by 
$m=p$ 
and $n=p-q$. 
The $L_\infty$ structure on $\scr B(\fg_{r+1})$ that we are going to introduce is based on yet another $\mathbb{Z}$-grading,
\begin{align}
\scr B(\fg_{r+1}) = \bigoplus_{\ell \in \mathbb{Z}} L_\ell\;,
\end{align}
where the degree $\ell$ of an element in $R_{(p,q)}$ is given by
$\ell=p+q$. The $L_\infty$ structure is then defined on (a part of)
the subalgebra of
$\scr B(\fg_{r+1})$ corresponding to positive levels $\ell$, and all
the brackets have level $\ell=-1$.
It is important, however, to note that the subset of $\scr
B(\fg_{r+1})$ on which the ghosts live is not closed under the
superalgebra bracket, so the space on which the $L_\infty$ algebra
is defined
will not support a Lie superalgebra structure.
The subset in question consists of the positive levels of the
subalgebra ${\scr B}(\fg_r)$ at $p>0$, $q=0$, together with a subset
of the elements at $p>0$, $q=1$.
See further Sections \ref{AncillaryGhostsSection} and
\ref{FullBVSection}.
The ghost number is identified with the level $\ell=p+q$ in Table
\ref{GeneralTable}.

\begin{table}[h]
\begin{align*}
\xymatrix@!0@C=2.2cm{
\cdots \ar@{-}[]+<1.1cm,1em>;[dddddd]+<1.1cm,-1em> \ar@{-}[]+<-0.8cm,-1em>;[rrrrrr]+<0.6cm,-1em>&
p=-1 \ar@{-}[]+<1.1cm,1em>;[dddddd]+<1.1cm,-1em>&
p=0 \ar@{-}[]+<1.1cm,1em>;[dddddd]+<1.1cm,-1em>&
p=1\ar@{-}[]+<1.1cm,1em>;[dddddd]+<1.1cm,-1em>&
p=2\ar@{-}[]+<1.1cm,1em>;[dddddd]+<1.1cm,-1em>&
p=3\ar@{-}[]+<1.1cm,1em>;[dddddd]+<1.1cm,-1em>&\cdots\\ \cdots&&&&&& *+[F-:red][red]{ n=0}\\
q=3 &&          &           &                    &     {{\ttR}}_3 \ar@{-}@[red][ur]& *+[F-:red][red]{ n=1}\\
q=2&&          &           &        {\tR_2} \ar@{-}@[red][ur]&        {\tR}_3 \oplus {{\ttR}}_3 \ar@{-}@[red][ur] & *+[F-:red][red]{ n=2 } \ar@{-}@[red][dl]\\
q=1&&{\bf 1} \ar@{-}@[red][dl]  &     R_1 \ar@{-}@[red][ur] &   R_2 \oplus \tR_2 \ar@{-}@[red][ur] &   {R}_3 \oplus {{\tR}}_3 \ar@{-}@[blue][ul] \ar@{-}@[blue][dr] & *+[F-:red][red]{ n=3 } \ar@{-}@[red][dl]\\
q=0&\overline R_1 &{\bf 1}\oplus{\bf adj}\oplus{\bf 1} \ar@{-}@[red][ur]  \ar@{-}@[red][dl]  & R_1 \ar@{-}@[blue][ul] \ar@{-}@[red][dl] \ar@{-}@[red][ur] & R_2 \ar@{-}@[blue]@[blue][ul]\ar@{-}@[red][ur]&  R_3 \ar@{-}@[blue][ul]& \cdots\\
\cdots\ar@{-}@[red][ur] &\overline R_1 & {\bf 1} & & *+[F-:blue][blue]{ \ell=1}\ar@{-}@[blue][ul] & *+[F-:blue][blue]{ \ell=2}*\frm{-}\ar@{-}@[blue][ul] & *+[F-:blue][blue]{ \ell=3}\ar@{-}@[blue][ul] 
}
\end{align*}\\
\caption{\it The general structure of the superalgebra
  ${\scr B}(\fg_{r+1})$.
  The blue lines are the $L_{\infty}$-levels, given by
  $\ell=p+q$. We also have $m=p$. Red lines are the usual levels in the level
  decomposition of ${\scr B}(\fg_{r+1})$, and form $\fg_{r+1}$ modules.
  Tables with specific examples are given in Section
  \ref{ExamplesSection}, and use the same gradings as this table.}
\label{GeneralTable}
\end{table}

Following ref. \cite{Palmkvist:2015dea}, we let
$E_M$ and $F^M$ be fermionic basis elements of $R_{(1,0)}=R_1$ and
$R_{(-1,0)}=\overline R_1$, respectively,
in the subalgebra ${\scr B}(\fg_r)$, while
 $\tE_M$ and $\tF^M$ are bosonic basis elements of $R_{(1,1)}=R_1$ and
$R_{(-1,-1)}=\overline R_1$ in the subalgebra $\fg_{r+1}$.
Furthermore, we let $T_\alpha$ be generators of $\fg_r$, and $(t_\alpha)_M{}^N$
representation matrices in the $R_1$ representation. Adjoint indices
will be raised and lowered with the Killing metric $\eta_{\alpha\beta}$
and its inverse.
Then the remaining (anti-)commutation relations of 
generators at levels $-1$, $0$ and $1$ in the
``local superalgebra'' (\ie, where also the right hand side
belongs to level $-1$, $0$ or $1$) that follow from the Chevalley--Serre relations are
\begin{align}
[T_\alpha,E_{M}]&=-(t_\alpha)_{M}{}^{N} E_{N}\;, &
[T_\alpha,\tE_{M}]&=-(t_\alpha)_{M}{}^{N}\tE_{N}\;,\nn\\ 
[k,E_{M}]&=-(\lambda,\lambda)E_{M}\;,  &
[\tk,\tE_{M}]&=(2-(\lambda,\lambda))\tE_{M}\;,\nn\\ 
[\tk,E_{N}]&=(1-(\lambda,\lambda))E_{N}\;, &
[k,\tE_{N}]&=(1-(\lambda,\lambda))\tE_{N}\;,\nn\\ 
[e,E_{N}]&=\tE_{N}\;, & [e,\tE_{N}]&=0\;,\nn\\
[f,E_{N}]&=0\;, & [f,\tE_{N}]&= E_{N}\;,
\label{borcherds-comm-rel}
\end{align}
\begin{align}
[T_\alpha,F^{N}]&=(t_\alpha)_{M}{}^{N} F^{M}\;, &
[T_\alpha,\tF^{N}]&=(t_\alpha)_{M}{}^{N}\tF^{M}\;,\nn\\ 
[k,F^{N}]&=(\lambda,\lambda)F^{N}\;, &
[\tk,\tF^{N}]&=((\lambda,\lambda)-2)\tF^{N}\;,\nn\\ 
[\tk,F^{N}]&=((\lambda,\lambda)-1)F^{N}\;, &
[k,\tF^{N}]&=((\lambda,\lambda)-1)\tF^{N}\;,\nn\\ 
[e,F^{N}]&=0\;, & [e,\tF^{N}]&=F^N\;,\nn\\
[f,F^{N}]&=-\tF^{N}\;, & [f,\tF^{N}]&= 0\;,
\end{align}
\begin{align}
[E_{M},F^{N}]&=-(t^\alpha)_{M}{}^{N} T_\alpha + \de_{M}{}^{N} k\;, &
[\tE_{M},\tF^{N}]&=-(t^\alpha)_{M}{}^{N} T_\alpha + \de_{M}{}^{N}
\tk\;, 
\nn\\
[E_{M},\tF^{N}]&=\de_{M}{}^{N} f\;, & [\tE_{M}, F^{N}]&=-\de_{M}{}^{N} e\;.
\end{align}
From this we get
\begin{align}
[[E_{M},F^{N}],E_{P}]&= f_{M}{}^{N}{}_{P}{}^{Q}E_{Q},	&
[[\tE_{M},\tF^{N}],\tE_{P}]&=
{\tf}_{M}{}^{N}{}_{P}{}^{Q} \tE_{Q}\;,				
\nn\\
[[E_{M},F^{N}],\tE_{P}]&= \de_{M}{}^{N}\tE_{P} + f_{M}{}^{N}{}_{P}{}^{Q}\tE_{Q}\;,
&
[[\tE_{M},\tF^{N}],E_{P}]&=\de_{M}{}^{N}  E_{P}+ f_{M}{}^{N}{}_{P}{}^{Q} E_{Q}\;,			
\nn\\
[[E_{M},\tF^{N}],E_{P}]&= 0,&
[[\tE_{M},F^{N}],\tE_{P}]&= 0\;,			
\nn\\
[[E_{M},\tF^{N}],\tE_{P}]&=\de_{M}{}^{N} E_{P},&
[[\tE_{M},F^{N}],E_{P}]&=  -\de_{M}{}^{N} \tE_{P}\;,
\label{delprodukter}
\end{align}
where
\begin{align}
f_{M}{}^{N}{}_{P}{}^{Q} &= (t_\alpha)_{M}{}^{N} (t^\alpha)_{P}{}^{Q}
- (\lambda,\lambda)\de_{M}{}^{N} \de_{P}{}^{Q}\;,
\end{align}
and
\begin{align}
\tf_{M}{}^{N}{}_{P}{}^{Q} &= (t_\alpha)_{M}{}^{N} (t^\alpha)_{P}{}^{Q}
+\big(2- (\lambda,\lambda)\big)\de_{M}{}^{N} \de_{P}{}^{Q}\;.
\end{align}
In particular we have the identities
\begin{align}
[[E_{M}, F^{N}],E_{P}]&=[[\tE_{M},\tF^{N}],E_{P}]+[[ E_{M},\tF^{N}],\tE_{P}]\;,\nn\\
[[\tE_{M}, \tF^{N}],\tE_{P}]&=[[E_{M},F^{N}],\tE_{P}]-[[ \tE_{M},F^{N}],E_{P}]\;,
\end{align}
which follow from acting with $e$ and $f$ on $[[E_{M}, \tF^{N}],E_{P}]=0$ and $[[\tE_{M}, F^{N}],\tE_{P}]=0$, respectively.

Continuing to level $2$, the generators $E_M$ and $\tE_M$ fulfil
certain ``covariantised Serre relations'', following from the Serre
relations for $e_0$ and $[e,e_0]$, the generators corresponding to the
roots $\beta_0$ and $\gamma_0$, respectively. The Serre relation in the
${\scr B}(\fg_r)$ subalgebra states that $[E_M,E_N]$ only
spans a submodule $R_2$ of the symmetric product of two $R_1$'s. The
complement of $R_2$ in the symmetric product
is $R(-2\lambda)$, the only module appearing in
the square of an object in a minimal orbit. Similarly,
the Serre relation in the $\fg_{r+1}$ subalgebra states that
$[\tE_M,\tE_N]$
only spans $\tR_2$, the complement of which is the highest module
in the antisymmetric product of two $R_1$'s.
The bracket $[E_M,\tE_N]$ spans $R_2\oplus\tR_2$.
The conjugate relations apply to $F^M$ and $\tF^M$.
We thus have
\begin{align}
[E_M,E_N]&\in R_2\;, & [F^M,F^N]&\in \overline{R_2}\;,\nn\\
[E_M,\tE_N]&\in R_2\oplus\tR_2\;,
          & [F^M,\tF^N]&\in \overline{R_2}\oplus\overline{\tR_2}\;,\nn\\
[\tE_M,\tE_N]&\in \tR_2\;, & [\tF^M,\tF^N]&\in \overline{\tR_2}\;. \label{LevelPlusMinusTwo}
\end{align}
The modules $\overline{R_2}$ and $\overline{\tR_2}$ are precisely the
ones appearing in the symmetric and antisymmetric parts of the section
constraint in Section \ref{SectionLieDetSection}.
For more details, \eg\ on the connection to minimal orbits and to a
denominator formula for the Borcherds superalgebra, we refer to
refs. \cite{Palmkvist:2015dea,Cederwall:2015oua,Cederwall:2017fjm}.
The (anti-)commutation relations with generators at level $\pm 1$ acting on those
in (\ref{LevelPlusMinusTwo}) at level $\mp 2$ follow from eqs. (\ref{delprodukter}) by the Jacobi identity.

An important property of $\sB(\fg_{r+1})$ is that any non-zero level
decomposes into doublets of the Heisenberg superalgebra spanned by
$e$, $f$ and
$h$. This follows from eqs. (\ref{superheisenberg}). An element at positive level and
height $0$ is annihilated by $\ad f$. It can be
``raised'' to height $1$ by $\ad e$ and lowered back by $\ad f$. We
define, for any element at a non-zero level $p$,
\begin{align}
  A^\sh&=
  \frac1p[A,e]\;,\\
  A^\fl&=-[A,f]\;. \label{musik}
\end{align}
Then $A=A^{\sh\fl}+A^{\fl\sh}$.
Occasionally, for convenience, we will write raising and lowering
operators acting on algebra elements. We then use the same symbols for
the operators: 
$\fl A=A^\fl$ and $\sh A=A^\sh$. 

As explained above $\scr B(\fg_{r+1})$ decomposes into $\fg_r$ modules, where 
we denote the one at level $p$ and height
$q$ by $R_{(p,q)}$.
Every $\fg_r$-module $R_p=R_{(p,0)}$ at level $p>0$ and height $0$ exists
also at height $1$. In addition there may be another module.
We write $R_{(p,1)}=R_p\oplus \tR_p$. Sometimes, $\tR_p$ may
vanish.
The occurrence of non-zero modules $\tR_p$ is responsible for the
appearance of ``ancillary ghosts''.\footnote{The notation $\widetilde R_p$ was used differently in ref. \cite{Palmkvist:2015dea}.
There, $\widetilde R_1, \widetilde R_2, \widetilde R_3, \ldots$ correspond to $R_1, \widetilde R_2, \ttR_3, \ldots$ here,
\ie, the representations on the diagonal $n=0$ in Table \ref{GeneralTable}. Thus it is only for $p=2$ that the meanings of the notation coincide.}

Let $A$ and $B$ be elements at positive level and height $0$ (or more generally,
annihilated by $\ad f$), and denote the total statistics of an element
$A$ by $|A|$.
The notation is such that $|A|$ takes the value $0$
for a totally bosonic element $A$ and $1$ for a totally fermionic
one. ``Totally'' means statistics of generators and components
together, so that a ghost $C$ always has $|C|=0$, while its derivative
(to be defined in eq. (\ref{der-def}) below) 
has $|dC|=1$.
This assignment is completely analogous to the assignment of statistics
to components in a superfield.
To be completely clear, our conventions are such that also fermionic
components and generators anticommute, so that if \eg\ $A=A^ME_M$ and $B=B^ME_M$ are
elements at level $1$ with $|A|=|B|=0$, then
$[A,B]=[A^ME_M,B^NE_N]=-A^MB^N[E_M,E_N]$.
A bosonic gauge parameter $A^M$ at level $1$ sits in an
element $A$ with $|A|=1$.

Some useful formulas involving raising and lowering operators are
easily derived:
\begin{align}
  [A,B^\sh]^\fl&=[A,B]\;,\\
  [A,B^\sh]^\sh&=-(-1)^{|B|}(\ad h)^{-1}[[h, A^\sh],B^\sh]\;.
  \label{movesharp}
\end{align}
Note that $[A^\sh,B^\sh]$ has height $2$ and lies in $\tR_{p_A+p_B}$, if
$p_A,p_B$ are the levels of $A,B$. The decomposition
\begin{align}
  [A,B^\sh]=[A,B]^\sh-(-1)^{|B|}(\ad h)^{-1}[[h, A^\sh],B^\sh]^\fl
\end{align}
provides projections of $R_{(p,1)}=R_p\oplus \tR_p$ on the two subspaces.

We will initially consider fields (ghosts) in the positive levels of
$\sB(\fg_r)$, embedded in $\sB(\fg_{r+1})$ at zero height. They can thus be
characterised as elements with positive (integer) eigenvalues of
$\ad h$ and zero eigenvalue of the adjoint action of the element $q$
in eq. (\ref{q-eigenvalue}).
Unless explicitly stated otherwise, elements in
$\sB(\fg_{r+1})$ will be ``bosonic'', in the sense that components
multiplying generators that are fermions will also be fermionic, as in
a superfield. This agrees with the statistics of ghosts. With such
conventions, the superalgebra
bracket $[\cdot,\cdot]$ is graded antisymmetric, $[C,C]=0$ when $|C|=0$.

\section{Section constraint and generalised Lie
  derivatives\label{SectionLieDetSection}} 

We will consider elements in certain subspaces of the algebra
${\scr B}(\fg_{r+1})$ which are also functions of coordinates transforming in $R_1=R(-\lambda)$, the
coordinates of an extended space.
The functional dependence is such that a (strong) section constraint
is satisfied. A derivative is in $\overline R_1=R(\lambda)$. Given the
commutation relations between $F^M$ and $\tF^M$ (which both provide
bases of $\overline R_1$), the section constraint can be expressed as
\begin{align}\label{SectionConstrains}
[F^M,F^N]\*_M\otimes\*_N&=0\;,\nn\\
[F^M,\tF^N]\*_M\otimes\*_N&=0\;,\nn\\
[\tF^M,\tF^N]\*_M\otimes\*_N&=0\;.
\end{align}
The first equation expresses the vanishing of $R_2$ in the symmetric
product of two derivatives (acting on the same or different fields),
the last one the vanishing of $\tR_2$ in the antisymmetric product,
and the second one contains both the symmetric and antisymmetric
constraint.
The first and third constraints come from the subalgebras ${\scr
  B}(\fg_r)$ and $\fg_{r+1}$, respectively, which gives a simple
motivation for the introduction of the double extension. By the Jacobi identity, they imply
\begin{align}\label{SectionConstrainsJacobi}
[[x,F^M],F^N]\*_{(M}\otimes\*_{N)}&=0\;,\nn\\
[[x,\tF^M],\tF^N]\*_{[M}\otimes\*_{N]}&=0\;
\end{align}
for any element $x \in {\scr B}(\fg_{r+1})$.
We refer to refs. \cite{Palmkvist:2015dea,Cederwall:2017fjm}
for details concerning \eg\ the importance of
eqs. \eqref{SectionConstrains} for the generalised Lie derivative, and
the construction of
solutions to the section constraint. 

The generalised Lie derivative, acting on an element in $R_1$, has the
form 
\begin{align}
  \LL_UV^M=U^N\*_NV^M+Z_{PQ}{}^{MN}\*_NU^PV^Q\;,
  \label{ZGenLieDer}
\end{align}
where the invariant tensor $Z$ has the universal expression
\cite{Bossard:2017aae,Cederwall:2017fjm}
\begin{align}
\sigma Z=-\eta_{\alpha\beta}t^\alpha\otimes t^\beta+(\lambda,\lambda)-1
\end{align}
($\sigma$ is the permutation operator),
\ie, $Z_{PQ}{}^{MN}=-\eta_{\alpha\beta}(t^\alpha)_P{}^N(t^\beta)_Q{}^M
+((\lambda,\lambda)-1)\delta_P^N\delta_Q^M$.
With the help of the structure constants of ${\scr B}(\fg_{r+1})$ it
can now be written \cite{Palmkvist:2015dea}
\begin{align} \label{bracketform}
\scr L_U V = [[U,\tF^{N}],\partial_{N}V^\sh]-[[\partial_{N}U^\sh,\tF^{N}],V]\;,
\end{align}
where $U=U^ME_M$, $V=V^ME_M$, with $U^M$ and $V^M$ bosonic.
The two terms in this expression corresponds to the first and second
terms in eq. \eqref{ZGenLieDer}, respectively, using the fourth and
seventh equations in \eqref{delprodukter}.
It becomes clear that the superalgebra ${\scr B}(\fg_r)$ does not
provide the structure needed to construct a generalised Lie
derivative,
but that ${\scr B}(\fg_{r+1})$ does.
In the following Section we will show that this construction not only
is made possible, but that the generalised Lie derivative arises
naturally from considering the properties of a derivative.

We introduce the following notation for the
antisymmetrisation, which will be the $2$-bracket in the $L_\infty$ algebra,
\begin{align}
2 \dlb U,V \drb = \scr L_U V  - \scr L_V U =
[[U,\tF^{N}],\partial_{N}V^\sh]
-[[\partial_{N}U^\sh,\tF^{N}],V] - (U \leftrightarrow V)\;.
\end{align}
For the symmetric part we have
\begin{align}
2 \llparenthesis U, V \rrparenthesis &= \scr L_U V  + \scr L_V U\nn\\&=[[U,\tF^{N}],\partial_{N}V^\sh]-[[\partial_{N}U^\sh,\tF^{N}],V]\nn\\
&\quad\;+[[V,\tF^{N}],\partial_{N}U^\sh]-[[\partial_{N}V^\sh,\tF^{N}],U]\nn\\
&=[[U,\partial_M V^\sh],\tF^M]-[[\partial_M U^\sh,V],\tF^M]\;,
\end{align}
where we have used the Jacobi identity.
If $\tR_2=0$, then
\begin{align}
[\tE_M, E_N]=[\tE_N,E_M]=-[E_M,\tE_N]
\end{align}
so that $[\partial_M \tU, V]=-[\partial_M U,\tV]$ and
$2 \llparenthesis U, V \rrparenthesis = \partial_M [[U,\tV],\tF^M]$.

In the cases where $\scr L_U \scr L_V - \scr L_U \scr L_V = \scr L_{\dlb U,V \drb}$ we get
\begin{align}
2 \llparenthesis \dlb U, V  \drb , W\rrparenthesis &= \scr L_{\dlb U,V \drb}W + \scr L_W \dlb U,V \drb\nn\\
&= 2\scr L_U \scr L_V W +\scr L_W \scr L_U V
= 3 \scr L_U \scr L_V W\nn\\
&= 3(2\scr L_U \scr L_V W -\scr L_W \scr L_U V )\nn\\
&= 3(\scr L_{\dlb U,V \drb}W - \scr L_W \dlb U,V \drb)=6 \dlb \dlb U, V  \drb , W\drb 
\end{align}
antisymmetrised in $U,V,W$.
These expressions, and their generalisations, will return with ghosts
as arguments in Section \ref{NonAncillaryBracketsSection}. Note however that $U$ and $V$ have bosonic
  components. They will be replaced by fermionic ghosts, which together
  with fermionic basis elements build bosonic elements. The bracket will
be graded symmetric.

\section{Derivatives, generalised Lie derivatives and other operators
\label{OperatorsSection}}

In this Section, we will start to examine operators on elements at
height $0$, which are functions of coordinates in $R_1$.
Beginning with a derivative, and attempting to get as close as
possible to a derivation property, we are naturally led to the
generalised Lie derivative, extended to all positive
levels.
The generalised Lie derivative is automatically associated with a
graded symmetry, as opposed to the graded antisymmetry of the algebra
bracket. This will serve as a starting point for the
$L_\infty$ brackets.
Other operators arise as obstructions to various desirable properties,
and will represent contributions from ancillary ghosts.
Various identities fulfilled by the operators will be derived; they
will all be essential to the formulation of the $L_\infty$ brackets
and the proof of
their identities.

\subsection{The derivative}

Define a derivative $d$: $R_{(p,0)}\rightarrow R_{(p-1,0)}$ ($p>0$) by
\begin{equation}
  dA=
  \left\{
  \begin{matrix}
    0\;, &  A\in R_{(1,0)}\;,\hfill\\
   [\*_MA^\sh,\tF^M]\;,& A\in R_{(p,0)}\;,\;p>1\;.
  \end{matrix}
  \right.
  \label{der-def}    
\end{equation}
It fulfils $d^2=0$
thanks to the section constraint.
At levels $p>1$ (and height $0$),
\begin{align}
  dA_p=\frac1p[\*_MA,F^M]\;.
  \label{DerivativepDep}
\end{align}
This follows from
\begin{align}
  [A_p^\sh,\tF^M]=\tfrac1p[[A_p,e],\tF^M]
  =\tfrac1p[A_p,F^M]+\tfrac1p[[A_p,\tF^M],e]\;,
\end{align}
where $[A_p,\tF^M]=0$ for $p>1$.

Only insisting on having a nilpotent derivative does not determine the
relative coefficients depending on the level $p$ in eq.
\eqref{DerivativepDep}. The subsequent considerations will however
depend crucially on the coefficient.

\subsection{Generalised Lie derivative from ``almost derivation''}

The derivative is not a derivation, but its failure to be one is of a
useful form. It consists of two parts, one being connected to the
generalised Lie derivative, and the other to the appearance of
modules $\tR_p$.
The almost-derivation property is derived using eq.
\eqref{movesharp}, which allows moving around raising operators at
the cost of introducing height $1$ elements.
Let $p_A,p_B$ be the levels of $A,B$.
One can then use the two alternative forms
\begin{align}
  [A,B]^\sh=\left\{
  \begin{matrix}
    [A,B^\sh]+(-1)^{|B|}\frac{p_A}{p_A+p_B}[A^\sh,B^\sh]^\fl\;,\\
    (-1)^{|B|}[A^\sh,B]-(-1)^{|B|}\frac{p_B}{p_A+p_B}[A^\sh,B^\sh]^\fl
  \end{matrix}
  \right.
\end{align}
to derive
\begin{align}\label{DerivationDerivation}
  d[A,B]&=[[A,\*_MB]^\sh,\tF^M]+[[\*_MA,B]^\sh,\tF^M]\nn\\
  &=[[A,\*_MB^\sh],\tF^M]
       +(-1)^{|B|}\frac{p_A}{p_A+p_B}[[A^\sh,\*_MB^\sh]^\fl,\tF^M]\nn\\
  &\qquad+(-1)^{|B|}[[\*_MA^\sh,B],\tF^M]
         -(-1)^{|B|}\frac{p_B}{p_A+p_B}[[\*_MA^\sh,B^\sh]^\fl,\tF^M]\nn\\
&=[[A,\tF^M],\*_MB^\sh]+[A,[\*_MB^\sh,\tF^M]]\nn\\
  &\qquad+(-1)^{|B|}[\*_MA^\sh,[B,\tF^M]]+(-1)^{|B|}[[\*_MA^\sh,\tF^M],B]\nn\\
  &\qquad+(-1)^{|B|}\frac{p_A\*_M^{(B)}-p_B\*_M^{(A)}}{p_A+p_B}
         [[A^\sh,B^\sh],\tF^M]^\fl \nn\\
&=[[A,\tF^M],\*_MB^\sh]+[A,dB]+\delta_{p_B,1}[A,[\*_MB^\sh,\tF^M]]\nn\\
         &\qquad+(-1)^{|B|}[\*_MA^\sh,[B,\tF^M]]
         +(-1)^{|B|}[dA,B]+\delta_{p_A,1}(-1)^{|B|}[[\*_MA^\sh,\tF^M],B]\nn\\
  &\qquad+(-1)^{|B|}\frac{p_A\*_M^{(B)}-p_B\*_M^{(A)}}{p_A+p_B}
         [[A^\sh,B^\sh],\tF^M]^\fl\nn\\
         &=[A,dB]+(-1)^{|B|}[dA,B]\nn\\
 &\qquad+\delta_{p_A,1}\left(
   [[A,\tF^M],\*_MB^\sh]+(-1)^{|B|}[[\*_MA^\sh,\tF^M],B]
   \right)\nn\\
 &\qquad-(-1)^{|A||B|}\delta_{p_B,1}\left(
   [[B,\tF^M],\*_MA^\sh]+(-1)^{|A|}[[\*_MB^\sh,\tF^M],A]
   \right)\nn\\
   &\qquad+(-1)^{|B|}\frac{p_A\*_M^{(B)}-p_B\*_M^{(A)}}{p_A+p_B}
         [[A^\sh,B^\sh],\tF^M]^\fl\;.
\end{align}
where superscript on derivatives indicate on which field they act.
We recognise the generalised Lie derivative from
eq. \eqref{bracketform} in the second and third lines in the last
step, and we define, for arbitrary $A,B$,
\begin{align}\label{GeneralGenLieDer}
\LL_AB=\delta_{p_A,1}\left(
   [[A,\tF^M],\*_MB^\sh]+(-1)^{|B|}[[\*_MA^\sh,\tF^M],B]\right)\;.
\end{align}
The extension is natural: a parameter $A$ with $p_A>1$ generates a
vanishing transformation, while the action on arbitrary elements is
the one which follows from demanding a Leibniz rule for the
generalised Lie derivative.
Note that bosonic components at level $1$ implies fermionic elements,
hence the signs in eqs. \eqref{bracketform} and \eqref{GeneralGenLieDer} agree.
The last term in eq. \eqref{DerivationDerivation} is present only if
$\tR_{p_A+p_B}$ is non-empty, since $[A^\sh,B^\sh]$ is an element at
height $2$ with $[A^\sh,B^\sh]^\sh=0$. We will refer to such terms as ancillary
terms, and denote them $-R^\fl(A,B)$, \ie,
\begin{align}
R^\fl(A,B)=-(-1)^{|B|}\frac{p_A\*_M^{(B)}-p_B\*_M^{(A)}}{p_A+p_B}
         [[A^\sh,B^\sh],\tF^M]^\fl\;.
\end{align}
A generic ancillary element will be an element $K^\fl\in R_p$ at height $0$
(or raised to $K$ at height $1$) obtained
from an element $B_M\in\tR_{p+1}$ at height $1$ as
$K^\fl=[B_M,\tF^M]$. The extra index
on $B_M$ is assumed to be ``in section''. See Section
\ref{AncillaryGhostsSection} for a more complete discussion.

The derivative is thus ``almost'' a derivation, but the derivation property is
broken by two types of terms, the generalised Lie derivative and an
ancillary term:
\begin{align}
  d[A,B]-[A,dB]-(-1)^{|B|}[dA,B]=\LL_AB-(-1)^{|A||B|}\LL_BA
  -R^\fl(A,B)\;.
  \label{AlmostDerivationProperty}
\end{align}
The relative factor with which the derivative acts on different levels
is fixed by the existence of the almost derivation property.

Eq. \eqref{AlmostDerivationProperty} states that the symmetry of $\LL_AB$ is
graded symmetric, modulo  terms with ``derivatives'', which in the end
will be associated with exact terms. This is good, since it means
that we, roughly speaking,
have gone from the graded antisymmetry of the superalgebra
bracket to the desired symmetry of an $L_\infty$ bracket.
The graded antisymmetric part of the generalised Lie derivative
appearing in eq. \eqref{AlmostDerivationProperty}
represents what, for bosonic parameters $U,V$, would be the symmetrised part
$\LL_UV+\LL_VU$, and it can be seen as responsible for the
violation of the Jacobi identities (antisymmetry and the Leibniz
property imply the Jacobi identities
\cite{Hitchin:2010qz}).
The
generalised Lie derivative (at level $1$) will be the starting point for
the $L_\infty$ $2$-bracket in Sections
\ref{NonAncillaryBracketsSection} and~\ref{FullBVSection}.

We note that $\LL_{dA}B=0$, $\LL_{[A,B]}C=0$, and that $\LL_A$ fulfils a Leibniz
rule,
\begin{align}
\LL_A[B,C]=(-1)^{|C|}[\LL_AB,C]+(-1)^{|A||B|}[B,\LL_AC]\;.
\end{align}

Consider the expression \eqref{GeneralGenLieDer} for the generalised
Lie derivative. It agrees with eq. \eqref{bracketform} when
$p_A=p_B=1$ and $|A|=|B|=1$.
It is straightforward to see that the expression contains a factor
$(-1)^{|B|+1}$ compared to the usual expression for the generalised
Lie derivative when expressed in terms of components.

In the present paper, we will assume that the generalised Lie derivative,
when acting on an element in ${\scr B}_+(\fg_r)$, close. This is {\it not}
encoded in the Borcherds superalgebra. We will indicate in the Conclusions
what we think will be the correct procedure if this is not the case.
We thus assume
\begin{align}
  (\LL_A\LL_B+(-1)^{|A||B|}\LL_B\LL_A)C
  =(-1)^{|C|+1}\LL_{\frac12(\LL_AB+(-1)^{|A||B|}\LL_BA)}C\;,
  \label{LLCommutator}
\end{align}
where the sign comes from the consideration above. When all components
are bosonic and level $1$, this becomes the usual expression
$(\LL_A\LL_B-\LL_B\LL_A)C=\LL_{\frac12(\LL_AB-\LL_BA)}C$. If we
instead consider a ghost $C$ with $|C|=0$, then
\begin{align}
\LL_C\LL_CC=-\tfrac12\LL_{\LL_CC}C\;.
\end{align}

\subsection{``Almost covariance'' and related
  operators\label{AlmostCovarianceSection}}  

The generalised Lie derivative anticommutes with the
derivative, modulo ancillary terms. This can be viewed as covariance
of the derivative, modulo ancillary terms. Namely, combining
eq. \eqref{AlmostDerivationProperty} with entries $A$ and $dB$ with
the derivative of eq. \eqref{AlmostDerivationProperty} gives the
relation
\begin{align}
  d\LL_AB+\LL_AdB&=(-1)^{|B|}([dA,dB]-d[dA,B])+(-1)^{|A||B|}d\LL_BA\nn\\
  &\quad\,+dR^\fl(A,B)+R^\fl(A,dB)\;.
\end{align}
The left hand side can only give a non-vanishing
contribution for $p_A=1$ and
$p_B>1$. But then the non-ancillary part of the right hand side
vanishes. Therefore, we can define an ancillary operator $X_AB$
as
\begin{align}
  d\LL_AB+\LL_AdB=-X^\fl_AB\;.
  \label{DefinitionOfX}
\end{align}
The explicit form of $X_A$ is
\begin{align}\label{ExplicitFormOfX}
  X^\fl_AB=-(d\LL_A+\LL_Ad)B=
  -\tfrac12\delta_{p_A,1}[[[\*_M\*_NA^\sh,B^\sh],\tF^M],\tF^N]\;.
\end{align}

The notation $X^\fl_AB$ means $(X_AB)^\fl$. Thus, $X_AB$ is an element
in $R_{p_B-1}$
at height $1$. It will be natural to extend the action of the derivative
and generalised Lie derivative to elements $K$ at height $1$ by
\begin{align}
  dK&=-(dK^\fl)^\sh\;,\nn\\
  \LL_CK&=-(\LL_CK^\fl)^\sh\;.
\end{align}
Then, $d\fl+\fl d=0$ and $\LL_C\fl+\fl\LL_C=0$.

Note that $X_{dA}B=0$ and $X_{[A,B]}C=0$, directly inherited from the
generalised Lie derivative.
In addition, we always have
\begin{align}
  \LL_{X^\fl_AB}C=0\,.
  \label{LLXIsZero}
\end{align}
If $\tR_2=0$ this statement is trivial. If $\tR_2$ is non-empty (as
\eg\ for $\fg_r=E_7$), $X^\fl_AB$ represents a parameter which gives a
trivial transformation without
being a total derivative, thanks to the section constraint.

\subsection{More operator identities}

The operator $X^\fl_A$ obeys the important property
\begin{align}
  dX^\fl_AB-X^\fl_AdB=0\;.
  \label{dXplusXd}
\end{align}
It follows from the definition of $X^\fl_A$ and the nilpotency of $d$ as
\begin{align}
  dX^\fl_AB-X^\fl_AdB=-d(d\LL_AB+\LL_AdB)+(d\LL_A+\LL_Ad)dB=0\;.
\end{align}
It can also be verified by the direct calculation
\begin{align}
dX^\fl_AB-X^\fl_AdB
  &=-\tfrac12\delta_{p_A,1}
  [[[\*_P[\*_M\*_NA^\sh,B^\sh],\tF^M],\tF^N]^\sh,\tF^P]^{\sh\fl}\nn\\
  &\quad\, 
  +\tfrac12\delta_{p_A,1}
  [[[\*_M\*_NA^\sh,[\*_PB^\sh,\tF^P]^\sh],\tF^M],\tF^N]^{\sh\fl}\nn\\
     &=\frac{\delta_{p_A,1}}{(p_B-1)(p_B-2)}\Bigl(
  [[[\*_P[\*_M\*_NA^\sh,B^\sh],\tF^M],F^N],F^P]\Bigr.\nn\\
  &\quad\, 
  \Bigl.+[[[\*_M\*_NA^\sh,[\*_PB^\sh,F^P]],\tF^M],F^N]
  \Bigr)\;,
\end{align}
where the action of the raising operators have been expanded.
In the first term, $\*_P$ must hit $B$, the other term vanishes due
to the section constraint. In the second term,
$[\*_M\*_NA^\sh,[\*_PB^\sh,F^P]]=[[\*_M\*_NA^\sh,\*_PB^\sh],F^P]$, and
the two terms cancel.
Note that we are now dealing with identities that hold exactly, not
only modulo ancillary terms (they are identities {\it for} ancillary terms).

An equivalent relation raised to height $1$ is
\begin{align}
  (dX_A+X_Ad)B=0\;.
  \label{dXPlusXd}
\end{align}
A relation for the
commutator of $X^\fl$ with $\LL$ is obtained directly from the
definition (\ref{DefinitionOfX}) of $X$,
\begin{align}
  &\left(\LL_AX^\fl_B-X^\fl_A\LL_B+(-1)^{|A||B|}(\LL_BX^\fl_A-X^\fl_B\LL_A)
  \right)C\nn\\
  &\qquad=(-1)^{|C|}X^\fl_{\frac12(\LL_AB+(-1)^{|A||B|}\LL_BA)}C\;,
\end{align}
or
\begin{align}
  &\left(\LL_AX_B+X_A\LL_B+(-1)^{|A||B|}(\LL_BX_A-X_B\LL_A)
  \right)C\nn\\
  &\qquad=(-1)^{|C|+1}X_{\frac12(\LL_AB+(-1)^{|A||B|}\LL_BA)}C\;.
\end{align}
For a ghost $C$ the relation reads
\begin{align}
  \LL_CX^\fl_CC-X^\fl_C\LL_CC=\tfrac12X^\fl_{\LL_CC}C\;,
  \label{LcXcCommutator}
\end{align}
or equivalently,
\begin{align}
  (\LL_CX_C+X_C\LL_C)C=-\tfrac12X_{\LL_CC}C\;.
  \label{XGenLieDerIdentity}
\end{align}

Further useful relations expressing derivation-like properties,
derived using the definitions of $X_AB$ and
$R(A,B)$, together with eq. \eqref{LLCommutator}, are:
\begin{align}
  dR(A,B)-R(A,dB)-(-1)^{|B|}R(dA,B)=X_AB-(-1)^{|A||B|}X_BA
  \label{RDerivationIdentity}
\end{align}
and
\begin{align}
  &\LL_AR(B,C)-(-1)^{|A||B|}R(B,\LL_AC)-(-1)^{|C|}R(\LL_AB,C)\nn\\
  &\qquad=-X_A[B,C]+(-1)^{|A||B|}[B,X_AC]+(-1)^{|C|}[X_AB,C]\;.
  \label{RGenLieDerIdentity}
\end{align}

Although $R(A,B)$ is non-vanishing for $A$ and $B$ at all levels 
(as long as $\tR_{p_A+p_B}$ is non-empty), we will sometimes use the
notation $R_AB=R(A,B)$.
Thanks to the Jacobi identity for the Borcherds superalgebra and the
Leibniz property of the generalised Lie derivative, $R(A,B)$ satisfies
a cyclic identity,
\begin{align}
  0&=R(A,[B,C])-R([A,B],C)-(-1)^{|A||B|}R(B,[A,C])\nn\\
  &\quad\,+[A,R(B,C)]-[R(A,B),C]-(-1)^{|A||B|}[B,R(A,C)]\;.
  \label{CyclicIdentity}
\end{align}

\section{Batalin--Vilkovisky ghost actions and
  \texorpdfstring{$L_\infty$}{Linfinity}
  algebras\label{BVLInftySection}} 

Let $C\in{\scr V}$ be a full set of ghosts, including ghosts for ghosts etc. If
the ``algebra'' of gauge transformations does not contain any field
dependence, the Batalin--Vilkovisky (BV) action \cite{Batalin:1981jr}
can be truncated to 
ghosts and their antifields $C^\star$. We denote this ghost action
$S(C,C^\star)$, and assume further that it is linear in $C^\star$.
The ghost action $S$ can be (formally, if needed) expanded as a power
series in $C$,
\begin{align}
S(C,C^\star)=\sum\limits_{n=1}^\infty\langle C^\star,\dlb C^n\drb\rangle\;,
\end{align}
where $\langle\cdot,\cdot\rangle$ is the natural scalar product on the
vector space of the ghosts and its dual, and where
\begin{align}
\dlb C^n\drb=\dlb\underbrace{C,C,\ldots,C}_n\drb
\end{align}
is a graded symmetric map from $\otimes^n{\scr V}$ to ${\scr V}$.
This map is, roughly speaking, the $L_\infty$ $n$-bracket. The $1$-bracket is the BRST operator.
The BV variation of $C$ is
\begin{align}
(S,C)=\sum\limits_{n=1}^\infty\dlb C^n\drb\;.
\end{align}
The BV master equation $(S,S)=0$ becomes, phrased as the nilpotency of
the transformation $(S,\cdot)$, the relation $(S,(S,C))=0$, which in
the series expansion turns into a set of identities for the brackets
\cite{Lada:1992wc,Zwiebach:1992ie,Hohm:2017pnh,Roytenberg:1998vn},
\begin{align}
  \sum\limits_{i=0}^{n-1}(i+1)\dlb C^i,\dlb C^{n-i}\drb\drb=0\;.
  \label{SSIdentities}
\end{align}

Often, $L_\infty$ algebras are presented with other conventions (see
ref. \cite{Hohm:2017pnh} for an overview). This
includes a shifted notion of level, equalling ghost number minus $1$.
Then the $n$-bracket carries level $n-2$. In our conventions, all
$L_\infty$
brackets carry ghost number $-1$, and the superalgebra bracket
preserves ghost number. Also, the properties of the
brackets under permutation of elements are sometimes presented as
governed by ``Koszul sign factors''. In our conventions, the
$L_\infty$ brackets
are simply graded symmetric and the statistics of the ghosts,
inherited from the superalgebra, is taking care of all signs automatically. 

Since the relation between the BV ghost variation and the $L_\infty$
brackets seems to be established, but not common knowledge among
mathematical physicists, we would like to demonstrate the equivalence
explicitly. 
(See also refs. \cite{Stasheff:1997iz,Hohm:2017pnh}.

In order to go from the compact form \eqref{SSIdentities} to a version
with $n$ arbitrary elements, let $C=\sum_{k=1}^\infty C_k$ and take the part
of the identity containing each of the terms in the sum once.
We then get
\begin{align}
  \sum\limits_{\substack{i,j\geq1\\ i+j=n+1}}j
  \sum\limits_\sigma\,\dlb C_{\sigma(i+1)},\ldots,C_{\sigma(n)},
  \dlb C_{\sigma(1)},\ldots,C_{\sigma(i)}\drb\drb=0\;,
  \label{IntermediateIdentity}
\end{align}
where the inner sum is over all permutations $\sigma$ of
$\{1,\ldots,n\}$.
The standard definition of the $L_\infty$ identities does not involve
the sum over all permutations, but over the subset of ``unshuffles'',
permutations which are ordered inside the two subsets:
\begin{align}
  \sigma(1)<\ldots<\sigma(i)\;,\nn\\
  \sigma(i+1)<\ldots<\sigma(n)\;.
\end{align}
Reexpressing the sum in terms of the sum over unshuffles gives a
factor $i!(n-i)!$, which combined with the factor $j$ in eq.
\eqref{IntermediateIdentity} gives $i!j!$,
Rescaling the brackets according to
\begin{align}
  n!\dlb C_1,\ldots,C_n\drb=\bar\ell(C_1,\ldots,C_n)
  \label{BracketRescaling}
\end{align}
turns the identity into
\begin{align}
  &\sum\limits_{\substack{i,j\geq1\\ i+j=n+1}}
  {\sum\limits_\sigma}'\,\bar\ell(C_{\sigma(1)},\ldots,C_{\sigma(j-1)},
  \bar\ell(C_{\sigma(j)},\ldots,C_{\sigma(n)}))=0\;,
\end{align}
where the primed inner sum denotes summation over unshuffles.

It remains to investigate the sign factors induced by the statistics
of the elements in the superalgebra.
We therefore introduce a basis $\{c_i\}$ which 
consists of fermionic elements with odd ghost numbers and bosonic elements with even ghost numbers. Since a ghost is always totally bosonic, this means that
ghosts with odd ghost numbers have fermionic components in this basis and ghosts with even ghost numbers have bosonic components.
Furthermore, we include the $x$-dependence of the ghosts in the basis elements $c_i$ (``DeWitt notation'') and thus treat the components as constants
that we can move out of the brackets.
Then, our identities take the form
\begin{align}
  \sum\limits_{\substack{i,j\geq1\\ i+j=n+1}}
             {\sum\limits_\sigma}'\,\varphi_{j-1}(\sigma;c)
             \bar\ell(c_{\sigma(1)},\ldots,c_{\sigma(j-1)},\bar\ell(
  c_{\sigma(j)},\ldots,c_{\sigma(n)}))=0\;,
  \label{Lbracketsc}
\end{align}
where $\varphi_{j-1}(\sigma;c)$ is the sign factor for the permutation
$\sigma$ in the graded
symmetrisation of the elements $\{c_1,\ldots,c_n,F\}$ to
$\{c_{\sigma(1)},\ldots,c_{\sigma(j-1)},F,c_{\sigma(j)},\ldots,c_{\sigma(n)}\}$. Here,
$F$ is a fermionic element used to define the sign factor, which
comes from the fact that the brackets are fermionic.

We now turn to the standard definition of $L_\infty$ identities.
The Koszul sign factor $\varepsilon(\sigma;x)$
for a permutation $\sigma$ of $n$ elements
$\{x_1,\ldots,x_n\}$ is defined inductively by an associative and
graded symmetric product 
\begin{align}
x_i\circ x_j=(-1)^{|x_i||x_j|}x_j\circ x_i\;,
\end{align}
where $|x_i|=0$ for ``bosonic'' $x_i$ and $1$ for
``fermionic''.
Then,
\begin{align}
  x_{\sigma(1)}\circ\ldots\circ x_{\sigma(n)}
  =\varepsilon(\sigma;x)\,x_1\circ\ldots\circ x_n\;.
\end{align}
Multiplying by a factor $(-1)^\sigma$ gives a graded antisymmetric
product, which can be seen as a wedge product of super-forms,
\begin{align}
  x_{\sigma(1)}\wedge\ldots\wedge x_{\sigma(n)}
  =(-1)^\sigma\varepsilon(\sigma;x)\,x_1\wedge\ldots\wedge x_n\;.
\end{align}

The standard form of the identities for an $L_\infty$ bracket is
\begin{align}
  &\sum\limits_{\substack{i,j\geq1\\ i+j=n+1}}(-1)^{i(j-1)}
             {\sum\limits_\sigma}'\,(-1)^\sigma\varepsilon(\sigma;x)
             \ell(\ell(x_{\sigma(1)},\ldots,x_{\sigma(i)}),
             x_{\sigma(i+1)},\ldots,x_{\sigma(n)}) =0\;.
 \label{Lbracketsx}
\end{align}

The two equations \eqref{Lbracketsc} and \eqref{Lbracketsx} look
almost identical. However,
the assignment of ``bosonic'' and ``fermionic'' for the $c$'s is
opposite to the one for the $x$'s.
On the other hand, the brackets of $x$'s are graded antisymmetric,
while those of $c$'s are graded symmetric. Seen as tensors, such
products differ in sign when exchanging bosonic with fermionic
indices.
There is obviously a difference between a tensor being graded
antisymmetric (the ``$x$ picture'') and ``graded symmetric with
opposite statistics'' (the ``$c$ picture'').
The two types of tensors are however equivalent as modules (super-plethysms) of
a general linear superalgebra.
As a simple example, a $2$-index tensor which is graded antisymmetric
can be represented as a matrix
\begin{align}
  \left(
  \begin{matrix}
    a&\alpha\cr
    -\alpha^t&s
  \end{matrix}
  \right)\;,
  \label{2Matrixx}
\end{align}
where $a$ is antisymmetric and $s$ symmetric,
while a $2$-index tensor which is graded symmetric in the opposite
statistics is 
\begin{align}
  \left(
  \begin{matrix}
    a'&\alpha'\cr
    (\alpha')^t&s'
  \end{matrix}
  \right)\;.
  \label{2Matrixc}
\end{align}
The tensor product $V\otimes V$ of a graded vector space $V$ with
itself can always be 
decomposed as the sum of the two plethysms, graded symmetric and
graded antisymmetric, \ie, in  the sum of the two
super-plethysms. Equivalently, the same decomposition, as modules of
the general linear superalgebra $\gl(V)$, is the sum of the graded
antisymmetric and graded symmetric modules with the opposite
assignment of statistics. The same is true for higher tensor products
$\otimes^nV$.

This means that, as long as the brackets $\ell$ and $\bar\ell$ are
taken to be proportional up to signs, the equations \eqref{Lbracketsc} and
\eqref{Lbracketsx} contain the same number of equations in the same
$\fg$-modules, but not that 
the signs for the different terms in the identities are equivalent.
In order to show this, one needs to
introduce an explicit invertible map, a so called suspension, from the ``$x$
picture'' to the ``$c$ picture'', \ie, between the two
presentations of the plethysms of the general linear superalgebra.

Let us use a basis where all basis elements are labelled by an index
$A=(a,\alpha)$, where $a$and and $\alpha$ correspond to
fermionic
and bosonic basis elements, respectively.
We choose an ordering where the $a$ indices are ``lower'' than
the $\alpha$ ones. Any unshuffle then has the index structure
$\{a_1\ldots a_k\alpha_1\ldots\alpha_{k'},
  a_{k+1}\ldots a_\ell\alpha_{k'+1}\ldots\alpha_{\ell'}\}$.
  If the brackets $\ell$ and
$\bar\ell$ are expressed in terms of structure constants,
\begin{align}
  \ell(x_{A_1},\ldots,x_{A_n})=f_{A_1\ldots A_n}{}^Bx_B\;,\nn\\
  \bar\ell(c_{A_1},\ldots,c_{A_n})=\bar f_{A_1\ldots A_n}{}^Bc_B\;,
\end{align}
the respective identities contain terms of the forms
\begin{align}
  &(-1)^{i(j-1)}(-1)^\sigma
  \varepsilon(a_1\ldots a_k\alpha_1\ldots\alpha_{k'}
     a_{k+1}\ldots a_\ell\alpha_{k'+1}\ldots\alpha_{\ell'})\nn\\
&\qquad\times  f_{a_1\ldots a_k\alpha_1\ldots\alpha_{k'}}{}^B
      f_{Ba_{k+1}\ldots a_\ell\alpha_{k'+1}\ldots\alpha_{\ell'}}{}^A\;,\nn\\
 &(-1)^m\varphi_{j-1}(a_{m+1}\ldots a_\ell\alpha_{m'+1}\ldots\alpha_{\ell'}
      a_1\ldots a_m\alpha_1\ldots\alpha_{m'})\label{ffbarfbarf}\\
 &\qquad\times\bar f_{a_{m+1}\ldots a_\ell\alpha_{m'+1}\ldots\alpha_{\ell'}}{}^B
      \bar f_{a_1\ldots a_m\alpha_1\ldots\alpha_{m'}B}{}^A\nn
      \;,
\end{align}
where $k+m=\ell$, $k'+m'=\ell'$, $k+k'=i$, $m+m'=j-1$ ($i,j$ being the
same variables as in the sums \eqref{Lbracketsc} and
\eqref{Lbracketsx}). Now, both expressions need to be arranged to the
same index structure, which  we choose as
$a_1\ldots a_\ell\alpha_1\ldots\alpha_{\ell'}$. 
This gives a factor $(-1)^{k'm}$ for the $f^2$ term, and $(-1)^{km}$
for $\bar f^2$. In order to compare the two brackets, we also need to
move the summation index $B$ to the right on $f$ when $B=\beta$ and to
the left on $\bar f$ when $B=b$. All non-vanishing brackets have a
total odd number of ``$a$ indices'', including the upper index,
so $B=b$ when $k$ is even, and $B=\beta$ when $k$ is odd. This gives a
factor $(-1)^m$ for the $f^2$ expression when $k$ is odd, and $(-1)^m$
for $\bar f^2$ when $k$ is even.

The task is now to find a relation
\begin{align}
  \bar f_{a_1\ldots a_k\alpha_1\ldots\alpha_{k'}}{}^B
  =\varrho(k,k')f_{a_1\ldots a_k\alpha_1\ldots\alpha_{k'}}{}^B
  \label{ffbarRelation}
\end{align}
for some sign $\varrho(k,k')$. The resulting relative sign
between the two expressions in eq. \eqref{ffbarfbarf} must then
be the same for all terms in an identity, \ie, it should only depend
on $\ell=k+m$ and $\ell'=k'+m'$. Taking the factors above into consideration,
this condition reads 
\begin{align}
  k\;\hbox{even}:
  &\quad(-1)^{(k+k')m'}\varrho(k,k')\varrho(m+1,m')=\tau(k+m,k'+m')\;,\nn\\
  k\;\hbox{odd}:
  &\quad(-1)^{(k+k')m'}\varrho(k,k')\varrho(m,m'+1)=\tau(k+m,k'+m')\;. 
\end{align}
This is satisfied for
\begin{align}
  \varrho(k,k')=(-1)^{\frac12k'(k'-1)}\;,
  \label{rhoExpression}
\end{align}
with $\tau(\ell,\ell')=\varrho(\ell,\ell')$. The last relation is
natural, considering that the equations in turn belong to the two
different presentations of the same super-plethysm.
This gives the explicit translation between the two pictures.

All structure constants carry an odd number of $a$ indices
(including the upper one). This is a direct consequence of the fact
that all brackets are fermionic in the $c$ picture (since the BV
antibracket is fermionic).
The relation between the structure constants in the two pictures
implies, among other things, that
\begin{align}  
  \bar f_a{}^\beta&=f_a{}^\beta\;,\nn\\
  \bar f_\alpha{}^b&=f_\alpha{}^b\;,\nn\\
  \bar f_{a_1a_2}{}^b&=f_{a_1a_2}{}^b\;,\label{ffbarExamples}\\
  \bar f_{a\alpha}{}^\beta&=f_{a\alpha}{}^\beta\;,\nn\\
  \bar f_{\alpha_1\alpha_2}{}^b&=-f_{\alpha_1\alpha_2}{}^b\;.\nn
\end{align}
The first two of these equations relate the $1$-bracket (derivative) 
in the two pictures, and the remaining three the $2$-bracket. Using these relations we can
give an explicit example of how identities in the two pictures are related to each other. Let us write $|c_a|=1$ and $|c_\alpha|=0$.
We then have
\begin{align}
\bar\ell(c_A,c_B)&=(-1)^{|c_A||c_B|}\bar\ell(c_B,c_A)\,,& \ell(x_A,x_B)&=-(-1)^{(|c_A|+1)(|c_B|+1)}\ell(x_B,x_A)\,.
\end{align}
Furthermore, the relations (\ref{ffbarExamples}) imply that under the inverse of the suspension,
\begin{align}
\bar\ell(c_A) &\mapsto \ell(x_A)\,,\nn\\
\bar\ell(c_A,c_B) &\mapsto (-1)^{|c_A|+1}\ell(x_A,x_B)\,. \label{desuspension}
\end{align}
In the $c$ picture, we have the identity
\begin{align} \label{c-identitet}
\bar\ell(\bar\ell(c_A,c_B))+(-1)^{|c_A|}\bar \ell(c_A,\bar\ell(c_B))+(-1)^{(|c_A|+1)|c_B|}\bar\ell(c_B,\bar\ell(c_A))=0\,.
\end{align}
Moving the inner $1$-bracket to the left, the left hand side is equal to the expression
\begin{align}
\bar\ell(\bar\ell(c_A,c_B)) 
+(-1)^{|c_A||c_B|}
\bar \ell(\bar\ell(c_B),c_A)+\bar\ell(\bar\ell(c_A),c_B)\,,
\end{align}
which, according to (\ref{desuspension}), is mapped to
\begin{align}
&(-1)^{|c_A|+1}\ell(\ell(x_A,x_B)) 
+(-1)^{(|c_A|+1)|c_B|}
\ell(\ell(x_B),x_A)+(-1)^{|c_A|}\ell(\ell(x_A),x_B)\\
&\qquad = (-1)^{|c_A|+1} \Big(  \ell(\ell(x_A,x_B)) 
+(-1)^{(|c_A|+1)(|c_B|+1)}
\ell(\ell(x_B),x_A)-\ell(\ell(x_A),x_B) \Big)\nn\\
&\qquad = (-1)^{|c_A|+1} \bigg(  \ell(\ell(x_A,x_B)) 
-\Big(\ell(\ell(x_A),x_B)-(-1)^{(|c_A|+1)(|c_B|+1)}
\ell(\ell(x_B),x_A)\Big) \bigg)\,.\nn
\end{align}
Setting this to zero 
gives the identity in the $x$ picture corresponding to the identity (\ref{c-identitet}) in the $c$ picture.

Note that the issue with the two pictures arises already when
constructing a BRST operator
in a situation where one has a mixture of bosonic and fermionic
constraints.
In the rest of the paper, we stay within the $c$ picture,
\ie, we work with ghosts with graded symmetry.

\section{The \texorpdfstring{$L_\infty$}{BV} structure, ignoring ancillary ghosts
  \label{NonAncillaryBracketsSection}}

The following calculation will first be performed disregarding
ancillary ghosts, \ie, as if all $\tR_p=0$. The results will form an
essential part
of the full picture, but the structure does not provide an $L_\infty$
subalgebra unless all $\tR_p=0$.

We use a ghost $C$ which is totally bosonic, \ie, $|C|=0$, and which is
a general element of ${\scr B}_+(\fg_r)$, \ie, a height $0$ element of
${\scr B}_+(\fg_{r+1})$. This gives
the correct statistics of the components, namely the same as the basis
elements in the superalgebra. All signs are taken care of
automatically by the statistics of the ghosts.
While the superalgebra bracket is graded antisymmetric, the
$L_\infty$ brackets (by which we mean the brackets in the $c$ picture
of the previous Section, before the rescaling of
eq. \eqref{BracketRescaling}) are graded symmetric.
The $a$ index of the previous Section labels ghosts with odd ghost
number, and the $\alpha$ index those with even ghost number, and
include also the coordinate dependence.

\subsection{Some low brackets}

The $1$-bracket acting on a ghosts at height $0$ is taken as
\begin{align}
  \dlb C\drb&=dC\;.
\end{align}
Then the $1$-bracket identity $\dlb\dlb C\drb\drb=0$ (the nilpotency of
the BRST operator) is satisfied.

The $2$-bracket on level $1$ elements $c$ is
\begin{align}
  \dlb c,c\drb=\LL_cc\;,
\end{align}
in order to reproduce the structure of the
generalised diffeomorphisms.
This already assumes that there are no ancillary
transformations, which also would appear on the right hand side of
this equation, and have their corresponding ghosts (we will comment on
this situation in the Conclusions).
It is natural to extend this to arbitrary levels by writing
\begin{align}
  \dlb C,C\drb=\LL_CC\;.
  \label{2bracket}
\end{align}

Given the relations \eqref{ffbarExamples}  between low brackets in the
two pictures in the previous Section, this essentially identifies the
1- and 2-brackets between components with the ones in the traditional
$L_\infty$ language (the $x$ picture).
Recall, however, that our ghosts $C$ are elements in the superalgebra,
formed as sums of components times basis elements, which lends a
compactness to the notation, which becomes  index-free.

There are potentially two infinities to deal with, one being the level
of the ghosts, the other the number of arguments in a bracket.
In
order to deal with the first one,
we are trying to derive a full set of $2$-brackets before going to higher
brackets. Of course, the existence of higher level ghosts is motivated
by the failure of higher identities, so it may seem premature to
postulate eq. \eqref{2bracket} before we have seen this
happen. However, it is essential for us to be able to deal with
brackets for arbitrary elements, without splitting them according to
level.
The identity for the $2$-bracket is then satisfied, since
\begin{align}
  \dlb\dlb C,C\drb\drb+2\dlb C,\dlb C\drb\drb
  =d\LL_CC+2\cdot\tfrac12\LL_CdC=0\;.
  \label{NonAnc2Bracket}
\end{align}
Notice that this implies that the $2$-bracket between ghosts which are
both at level $2$ or higher vanishes.

There is of course a choice involved every time a new bracket is
introduced, and the choices differ by something exact. The choice will
then have repercussions for the rest of the structure. The first
choice arises when the need for a level $2$ ghost $C_2$ becomes clear (from
the $3$-bracket identity as a modification of the Jacobi identity), and its
$2$-bracket with the level $1$ ghost is to be determined. Instead of
choosing $\dlb c,C_2\drb=\tfrac12\LL_cC_2$, corresponding to
eq. \eqref{NonAnc2Bracket}, we could have taken
$\dlb c,C_2\drb=-\tfrac12[c,dC_2]$, since the derivative of the two
expressions are the same (modulo ancillary terms) according to
eq. \eqref{AlmostDerivationProperty}. The latter is the type of choice made
in \eg\ ref. \cite{Hohm:2017pnh}. Any linear combination of the two choices
with weight $1$ is of course also a solution. However, it turns out that
other choices than the one made here lead to expressions that do not
lend themselves to unified expressions containing $C$ as a generic
element in ${\scr B}_+(\fg_r)$.
Thus, this initial choice and its continuation are of importance.

We now turn to the $3$-bracket.
The identity is
\begin{align}
  \dlb\dlb C,C,C\drb\drb+2\dlb C,\dlb C,C\drb\drb
  +3\dlb C,C,\dlb C\drb\drb=0\;.
  \label{C1C1C1relation}
  \end{align}
The second term (the Jacobiator) equals
$\LL_C\dlb C,C\drb+\LL_{\dlb C,C\drb}C$. Here we must
assume the closure of the transformations, acting on something, \ie,
the absence of ancillary transformations in the commutator of two
level $1$ transformations. Then,
\begin{align}
\LL_C\dlb C,C\drb=\LL_C\LL_CC=-\tfrac12\LL_{\dlb C,C\drb}C\;,
\end{align}
and the second term
in eq. \eqref{C1C1C1relation} can be written expressed in terms of the
(graded) antisymmetric part instead of the symmetric one, so that the
derivation property may be used:
\begin{align}
  2\dlb C,\dlb C,C\drb\drb&=-\tfrac13(\LL_C\LL_CC-\LL_{\LL_CC}C)\nn\\
    &=-\tfrac13\left(
d[C,\LL_CC]-[C,d\LL_CC]+[dC,\LL_CC]
\right)\nn\\
&=-\tfrac13\left(
d[C,\LL_CC]+[C,\LL_CdC]+[dC,\LL_CC]\right)
\end{align}
(modulo ancillary terms). If one takes
\begin{align}
\dlb C,C,C\drb=\tfrac13[C,\LL_CC]\;,
\end{align}
the identity is satisfied, since then
\begin{align}
\dlb\dlb C,C,C\drb\drb=\tfrac13d[C,\LL_C]\;,
\end{align}
and
\begin{align}
  3\dlb C,C,\dlb C\drb\drb
  =3\cdot\tfrac13(\tfrac13[C,\LL_CdC]+\tfrac13[dC,\LL_CC])\;.
\end{align}

Starting from the $4$-bracket identity
\begin{align}
  \dlb\dlb C,C,C,C\drb\drb
  +2\dlb C,\dlb C,C,C\drb\drb
  +3\dlb C,C,\dlb C,C\drb\drb+4\dlb C,C,C,\dlb C\drb\drb=0\;,
  \label{cccc4bracket}
\end{align}
a calculation gives at hand that the second and third terms
cancel (still modulo ancillary terms). This would allow
$\dlb C,C,C,C\drb=0$.
The calculation goes as follows.
We use the brackets and identities above to show
\begin{align}
  \dlb C,\dlb C,C,C\drb\drb&=\tfrac13\dlb C,[C,\LL_CC]\drb
  =\tfrac16\LL_C[C,\LL_CC]\nn\\
  &=-\tfrac16[\LL_CC,\LL_CC]+\tfrac16[C,\LL_C\LL_CC]\nn\\
  &=-\tfrac16[\LL_CC,\LL_CC]-\tfrac1{12}[C,\LL_{\LL_CC}C]
\end{align}
and
\begin{align}
  \dlb C,C,\dlb C,C\drb\drb
  &=\tfrac19\left([\LL_CC,\LL_CC]+[C,\LL_{\LL_CC}C]+[C,\LL_C\LL_CC]\right)\nn\\
  &=\tfrac19[\LL_CC,\LL_CC]+\tfrac1{18}[C,\LL_{\LL_CC}C]\;.
\end{align}

This does not imply that all higher brackets vanish. Especially, the
middle term $3\dlb C,C,\dlb C,C,C\drb\drb$ in the $5$-bracket identity
is non-zero, which requires
a $5$-bracket.

\subsection{Higher brackets
\label{HigherNonAncillaryBracketsSection}}

In order to go further, we need to perform calculations at arbitrary
order. There is essentially one possible form for the $n$-bracket, namely
\begin{align}
  \dlb C^n\drb=k_n(\ad C)^{n-2}\LL_CC\;.
\end{align}
It turns out that the constants $k_n$ are given by Bernoulli numbers,
\begin{align}
  k_{n+1}=\frac{2^nB^+_n}{n!}\;,
  \label{kInTermsOfBernoulli}
\end{align}
where $B^+_n=(-1)^nB_n$ (which only changes the sign for $n=1$, since
higher odd Bernoulli numbers are $0$).

We will first show that it is consistent to set all
$\dlb C^{2n}\drb=0$, $n\geq2$.
Then the $2(n+1)$-identity reduces to
\begin{align}
  0=2\dlb C,\dlb C^{2n+1}\drb\drb+(2n+1)\dlb C^{2n},\dlb C,C\drb\drb
  \;.
  \label{ZeroEvenBrackets}
\end{align}
Evaluating the two terms gives
\begin{align}    \label{EvenBracketZeroCalcI}
  \dlb C,\dlb C^{2n+1}\drb\drb
  &=\dlb C,k_{2n+1}(\ad C)^{2n-1}\LL_CC\drb\nn\\
  &\hskip-8mm=\tfrac12k_{2n+1}\LL_C(\ad C)^{2n-1}\LL_CC\\
  &\hskip-8mm=\tfrac12k_{2n+1}\Bigl((\ad C)^{2n-1}\LL_C\LL_CC
   -\sum\limits_{i=0}^{2n-2}(\ad C)^i\ad\LL_CC(\ad
   C)^{2n-2-i}\LL_CC\Bigr)\nn\\
  &\hskip-8mm=\tfrac12k_{2n+1}\Bigl(-\tfrac12(\ad C)^{2n-1}\LL_{\LL_CC}C
   -\sum\limits_{i=0}^{2n-2}(\ad C)^i\ad\LL_CC(\ad
   C)^{2n-2-i}\LL_CC\Bigr)\;,\nn
\end{align}
\begin{align}\label{EvenBracketZeroCalcII}
   \dlb C^{2n},\dlb C,C\drb\drb
   &=\frac{k_{2n+1}}{2n+1}
   \Bigl((\ad C)^{2n-1}\LL_C\LL_CC+(\ad C)^{2n-1}\LL_{\LL_CC}C\Bigr.\nn\\
   &\qquad\qquad\qquad\Bigl.
   +\sum\limits_{i=0}^{2n-2}(\ad C)^i\ad\LL_CC(\ad C)^{2n-2-i}\LL_CC\Bigr)\\
   &\hskip-8mm=\frac{k_{2n+1}}{2n+1}
   \Bigl(\tfrac12(\ad C)^{2n-1}\LL_{\LL_CC}C
    +\sum\limits_{i=0}^{2n-2}(\ad C)^i\ad\LL_CC(\ad
    C)^{2n-2-i}\LL_CC\Bigr)\;,\nn
\end{align}
which shows that eq. \eqref{ZeroEvenBrackets} is fulfilled.

We then turn to the general $n$-identities, $n\geq2$ (the remaining
ones are those with odd $n$). They are
\begin{align}
  0&=\dlb\dlb C^n\drb\drb
  +\sum\limits_{i=1}^{n-2}(i+1)\dlb C^{i},\dlb C^{n-i}\drb\drb
  +n\dlb C^{n-1},\dlb C\drb\drb\;.
  \label{OddcracketIdentity}
\end{align}
The first term equals
$k_nd(\ad C)^{n-2}\LL_CC$. Repeated use of
eq. \eqref{AlmostDerivationProperty} (without the ancillary term) gives
\begin{align}
  &d(\ad C)^{n-2}\LL_CC=-\sum\limits_{i=0}^{n-3}
        (\ad C)^i\ad dC(\ad C)^{n-i-3}\LL_CC\nn\\
  &\qquad-\tfrac n2 (\ad C)^{n-3}\LL_{\LL_CC}C
  -\sum\limits_{i=0}^{n-4}(i+1)(\ad C)^i\ad\LL_CC(\ad
  C)^{n-i-4}\LL_CC\;.
  \label{FirstAndLastTerms}
\end{align}
The first sum cancels the last term in eq. \eqref{OddcracketIdentity}.
We now evaluate the middle terms under the summation sign in
eq. \eqref{OddcracketIdentity}.
\begin{align}\label{MiddleTerms}
  \dlb C^{i},\dlb C^{n-i}\drb\drb
  &=\frac{k_{i+1}k_{n-i}}{i+1}
  \Bigl(-\tfrac12(\ad C)^{n-3}\LL_{\LL_CC}C\Bigr.\nn\\
  &\qquad+\sum\limits_{j=0}^{i-2}(\ad C)^j\ad((\ad C)^{n-i-2}\LL_CC)(\ad
            C)^{i-j-2}\LL_CC\nn\\
  &\qquad-\sum\limits_{j=0}^{n-i-3}(\ad C)^{i+j-1}\ad \LL_CC(\ad
            C)^{n-i-j-3}\LL_CC
           \Bigl.\Bigr).
\end{align}
Here we have ignored the insertion of the $2$-bracket in the argument of
the generalised Lie derivative in the $(n-1)$-bracket (which changes the
sign of the term with $\LL_{\LL_CC}C$), since this already has been
taken care of in eqs. \eqref{EvenBracketZeroCalcI} and
\eqref{EvenBracketZeroCalcII}. It does not appear 
in the identity for odd $n$.

Let $n=2m+1$ and $i=2j$.
There is a single
term containing $\LL_{\LL_CC}C$, namely
\begin{align}
-\frac{k_{2j+1}k_{2(m-j)+1}}{2(2j+1)}(\ad C)^{2m-2}\LL_{\LL_CC}C\;.
\end{align}
The total coefficient of this term in eq. \eqref{OddcracketIdentity}
demands that
\begin{align}
  k_{2n+1}=-\tfrac1{2n+1}\sum\limits_{j=1}^{n-1}k_{2j+1}k_{2(n-j)+1}\;.
  \label{FirstCoeffRelation}
\end{align}
It is straightforward to show that the Bernoulli numbers satisfy the identity
\begin{align}
  \sum\limits_{j=1}^{m-1}\frac{B_{2j}B_{2(m-j)}}{(2j)!(2(m-j))!}
  =-(2m+1)\frac{B_{2m}}{(2m)!}\;.
\end{align}
It follows from the differential equation
$\frac d{dt}[t(f-\frac{t^2}{12})]+f^2=0$, satisfied by
\begin{align}
f(t)=\frac t{e^t-1}+\frac t2-1\,,\qquad \text{where} \qquad \frac t{e^t-1}=\sum_{n=0}^\infty \frac{B_n}{n!}t^n\,.
\end{align}
The $(2m+1)$-identity \eqref{OddcracketIdentity} then is satisfied
with the coefficients given by eq. \eqref{kInTermsOfBernoulli}.
The initial value $k_3=\tfrac13$
fixes the coefficients to the values in
eq. \eqref{kInTermsOfBernoulli}.
Bernoulli numbers as coefficients of $L_\infty$ brackets have been
encountered earlier \cite{Bering:2006eb,Getzler:2010}.

In order to show that the identities are satisfied at all levels, we
use the method devised by Getzler \cite{Getzler:2010}
(although our expressions seem to be quite different from the ones in
that paper). 
All expressions remaining after using the derivation property
and identifying the coefficients using the $\LL_{\LL_CC}C$ terms are of
the form
\begin{align}
Z_{n,j,k}=(\ad C)^{n-4-j-k}[(\ad C)^j\LL_CC,(\ad C)^k\LL_CC]\;.
\end{align}
There are however many dependencies among these expressions.
First one observes that, since $\LL_CC$ is fermionic,
$Z_{n,j,k}=Z_{n,k,j}$. Furthermore, the Jacobi identity immediately
gives
\begin{align}
  Z_{n,j,k}=Z_{n,j+1,k}+Z_{n,j,k+1}
  \label{ZRelation}
\end{align}
for $j+k<n-4$.
If one associates the term $Z_{n,j,k}$ with  the monomial $s^jt^k$,
the Jacobi identity implies $s^jt^k\approx s^{j+1}t^k+s^jt^{k+1}$,
\ie, $(s+t-1)s^jt^k\approx0$. We can then replace $s$ by $1-t$, so
that $s^jt^k$ becomes $(1-t)^jt^k$. The symmetry property is
taken care of by symmetrisation, so that the final expression
corresponding to
$Z_{n,j,k}$ is
\begin{align}
\tfrac12((1-t)^jt^k+t^j(1-t)^k)\,.
\end{align}
All expressions are reduced to polynomials of degree up to $n-4$ in
one variable, symmetric under $t\leftrightarrow1-t$. An independent
basis consists of even powers of $t-\tfrac12$. In addition to the
equations with $\LL_{\LL_CC}C$ that we have already checked, there
are $m-1$ independent equations from the terms with $(\LL_CC)^2$
in the $(2m+1)$-identity, involving $k_{2m+1}$ and products of
lower odd $k$'s. 

We will now show that all identities are satisfied by translating them
into polynomials with Getzler's method, using the generating function
for the Bernoulli numbers.

Take the last sum in eq. \eqref{FirstAndLastTerms}. It represents the
contribution from the first and last terms in the identity.
It translates into the polynomial
\begin{align}
  -k_n\sum\limits_{i=0}^{n-4}(i+1)t^{n-i-4}
  =-k_n\frac{n-3-(n-2)t+t^{n-2}}{(1-t)^2}\;.
  \label{KnPolynomial}
\end{align}
The terms from the middle terms in the identity
(eq. \eqref{MiddleTerms})
translate into
\begin{align}
  &\sum\limits_{i=1}^{n-2}
  k_{i+1}k_{n-i}\Bigl(\sum\limits_{j=0}^{i-2}s^{n-i-2}t^{i-j-2}
  -\sum\limits_{j=0}^{n-i-3}t^{n-i-j-3}\Bigr)\nn\\
  &\quad=\sum\limits_{i=1}^{n-2}k_{i+1}k_{n-i}\Bigl(s^{n-i-2}\frac{1-t^{i-1}}{1-t}
  -\frac{1-t^{n-i-2}}{1-t}\Bigr)\;.
  \label{MiddlePolynomials}
\end{align}

Let $f(x)$ be the generating function for the coefficients $k_n$,
\ie,
\begin{align}
  f(x)&=\sum\limits_{n=2}^\infty k_nx^n
  =\sum\limits_{n=1}^\infty\frac{2^nB^+_n}{n!}x^{n+1}
  =\frac{2x^2}{1-e^{-2x}}-x\nn\\
  &=x^2+\tfrac13x^3-\tfrac1{45}x^5+\tfrac2{945}x^7-\tfrac1{4\,725}x^9
  +\tfrac2{93\,555}x^{11}-\tfrac{1\,382}{638\,512\,875}x^{13}+\cdots
\end{align}
We now multiply the contributions from eqs. \eqref{KnPolynomial}
and \eqref{MiddlePolynomials}, symmetrised in $s$ and $t$,
by $x^n$ and sum over $n$, identifying
the function $f$ when possibility is given.
This gives
\begin{align}
  &\frac1{2(1-t)^2}\Bigl(-(1-t)xf'(x)+(3-2t)f(x)-\frac{f(tx)}{t^2}
  \Bigr)\nn\\
  &+\frac1{2(1-t)x}\Bigl(
  -f(x)^2+\frac{f(x)f(sx)}{s^2}+\frac{f(x)f(tx)}{t^2}
  -\frac{f(sx)f(tx)}{s^2t^2}
  \Bigr)+(s\leftrightarrow t)\;.
\end{align}
When the specific function $f$ is used, this becomes, after some
manipulation,
\begin{align}
  \phi(s,t,x)=\frac{(s+t-1)x}{2st}
  &-\frac{(s+t-1)(2-s)x^2}{2t(1-s)^2}\frac{\sinh((1-s)x)}{\sinh
    x\sinh(sx)}
  \nn\\
  &-\frac{(s+t-1)(2-t)x^2}{2s(1-t)^2}\frac{\sinh((1-t)x)}{\sinh
    x\sinh(tx)}\nn\\
  &\quad\hskip-24mm+\frac{(s+t-2)x^3}{2(1-s)(1-t)}\frac1{\sinh^2x}
  \Bigl(1-\frac{\sinh((1-s)x)\sinh((1-t)x)}{\sinh(sx)\sinh(tx)}\Bigr)\;.
  \label{Phistx}
\end{align}
This expression clearly vanishes when $s+t-1=0$,
which proves that the identities
for the brackets hold to all orders.

The function $\phi(s,t,x)=\sum_{n=2}^\infty\phi_n(s,t)$, with the
coefficient functions $\phi_n(s,t)$ given by the sum of
eqs. \eqref{KnPolynomial} and \eqref{MiddlePolynomials}, symmetrised
in $s$ and $t$, will appear again in many of the calculations for the
full identities in Section \ref{FullBVSection}.

The complete variation $(S,C)=\sum_{n=1}^\infty\dlb C^n\drb$ can
formally be written as
\begin{align}
(S,C)=dC+g(\ad C)\LL_CC\;,
\end{align}
where $g$ is the function
\begin{align}
g(x)=\frac1{x^2}f(x)=\frac2{1-e^{-2x}}-\frac1x\;.
\end{align}
Then $(S,(S,C))=0$.
This concludes the analysis in the absence of ancillary terms.

\section{Ancillary ghosts\label{AncillaryGhostsSection}}

We have already encountered ``ancillary terms'', whose appearance in
various identities for the operators, such as the deviation of $d$ from
being a derivation and the deviation of $d$ from being covariant, rely
on the existence of modules $\tR_p$. Note that the Borcherds
superalgebra always has $\tR_1=\emptyset$, \ie, $R_{(1,1)}=R_1$; this is what prevents us
from treating situations where already the gauge ``algebra'' of
generalised Lie derivatives contains ancillary transformations.
The ancillary terms at level $p$
appear as $[B_M^\sh,\tF^M]^\fl=[B_M,\tF^M]$‚ where
$B_M$ is an element in $\tR_{p+1}$ at height $1$ (\ie, $B_M^\fl=0$).
$B_M$ carries an extra $\overline R_1$ index, which is ``in section'',
meaning that the relations \eqref{SectionConstrains} are fulfilled
also when one or two $\*_M$'s are replaced by a $B_M$.

The appearance of ancillary terms necessitates the introduction of
ancillary ghosts. We will take them as elements $K_p\in R_p$ at height $1$
constructed as above. The idea is then to extend the $1$-bracket to
include the operator $\fl$, which makes it possible to cancel
ancillary terms in identities (ignored in the previous Section) by a
``derivative'' $\fl$ of other terms at height $1$.

The derivative $d$ and the generalised Lie derivative $\LL_C$ are
extended to level $1$ as in Section
\ref{AlmostCovarianceSection}. This implies that $\fl$ anticommutes
with $d$ and with $\LL_C$. Since $d^2=0$ and $d\fl+\fl d=0$ on elements
in $R_p$ at height $0$ and $1$, it can be used in the construction of
a $1$-bracket, including the ancillary ghosts. The generic structure
is shown in Table \ref{DerivativeActionTable}.

Ancillary elements form an ideal ${\scr A}$ of ${\scr B}_+(\fg_r)$. Let
$K^\fl=[B_M,\tF^M]$ as above, and let $A\in{\scr B}_+(\fg_r)$. Then,
\begin{align}
[A,K^\fl]=[[A,B_M],\tF^M]+(-1)^{|A||B|}[B_M,[A,\tF^M]]\;.
\end{align}
The first term is ancillary, since the height $1$ element
$[A,B_M]$ is an element in
$\tR_{p_A+p_B}$, thanks to $[A,B_M]^\fl=0$, and the section property
of the $M$ index remains. The second term has $[A,\tF^M]\neq0$ only
for $p_A=1$‚ but vanishes thanks to $[B_M,f]=0$.
This shows that $[{\scr B}_+(\fg_r),{\scr A}]\subset{\scr A}$.
An explicit example of this ideal, for
the $E_5$ exceptional field theory in the
M-theory section, is given in Section
\ref{ExamplesSection}, Table \ref{ExceptionalDectionDiffTable}. 

Let us consider the action of $d$ on ancillary ghosts $K$ at height
$1$. Let $B_M\in\tR_{p+1}$ with height $1$, and let $K^\fl=[B^\sh_M,\tF^M]^\fl\in
R_p$ at height $0$. We will for the moment assume that
\begin{align}\label{BshWithF}
  [B^\sh_M,F^M]=0\;.
\end{align}
This is a purely algebraic condition stating that
$R_p$ but not $\tR_p$ is present in the tensor product
$\tR_{p+1}\otimes\overline R_1$ in $B_M$. 
Then, $K=[B^\sh_M,\tF^M]$. Acting with the derivative gives
\begin{align}
  dK=\tfrac1{p-1}[[\*_NB^\sh_M,\tF^M],F^N]
  =\tfrac1{p-1}[[\*_NB^\sh_M,F^N],\tF^M]=[B'^\sh_M,\tF^M]\;,
\end{align}
where $B'^\sh_M=\frac1{p-1}[\*_NB^\sh_M,F^N]$. The derivative
preserves the structure, thanks to the section constraint. Also, the
condition \eqref{BshWithF} for $B'$, $[B'^\sh_M,F^M]=0$, is
automatically satisfied. 

The appearance of modules $\tR_p$ can be interpreted in several
ways. One is as a violation of covariance of the exterior derivative,
as above. Another is as a signal that Poincar\'e's lemma does not
hold.
In this sense, ancillary modules encode the presence of ``local
cohomology'', \ie, cohomology present in an open set. It will be
necessary to introduce ghosts removing
this cohomology.

Let the lowest level $p$ for which $\tR_{p+1}$ is non-empty be $p_0$.
Then it follows that an ancillary element $K_{p_0}$ at level $p_0$
will be closed, 
$dK_{p_0}=0$, and consequently $dK^\fl_{p_0}=0$. However, $K_{p_0}$
does not need to be a total derivative, since $B_M$ does not need to
equal $\*_M\Lambda$. Indeed, our ancillary terms are generically not
total derivatives. An ancillary element at level $p_0$ represents a
local cohomology, a violation of Poincar\'e's lemma.

\begin{table}
\begin{align*}
  \xymatrix{
    &&&&K_{p_0}\ar[d]_\fl\ar@{<-}[r]_d
    &K_{p_0+1}\ar[d]_\fl\ar@{<-}[r]_d
    &K_{p_0+2}\ar[d]_\fl\ar@{<-}[r]_d&\cdots\\
    0\ar@{<-}[r]_d
    &C_1\ar@{<-}[r]_d
    &\cdots\ar@{<-}[r]_d
    &C_{p_0-1}\ar@{<-}[r]_d
    &C_{p_0}\ar@{<-}[r]_d
    &C_{p_0+1}\ar@{<-}[r]_d
    &C_{p_0+2}\ar@{<-}[r]_d&\cdots\\
}
\end{align*}
\caption{\it The typical structure of the action of the $1$-bracket
  between the ghost modules,
  with ancillary ghosts appearing from level $p_0\geq1$.}
\label{DerivativeActionTable}
\end{table}

The algebraic condition \eqref{BshWithF} was used to show that the
ancillary property is preserved under the derivative.
Consider the expression $X_AB$ from
eq. \eqref{ExplicitFormOfX}. Raised to height $1$ it gives an
expression
\begin{align}
K=[[\beta_{MN},F^N],\tF^M]=[B_M^\sh,\tF^M]
\end{align}
with $B^\sh_M=[\beta_{MN},F^N]$, where $\beta_{MN}$ is symmetric and where
both its indices are in section. 
Then, $[B^\sh_M,F^M]=0$, and the condition is satisfied.
The same statement can not be made directly for any term $R(A,B)$,
since it contains only one derivative. 
One can however rely the identities
\eqref{RDerivationIdentity} and
\eqref{RGenLieDerIdentity}, which immediately show (in the latter case also
using the property that ancillary expressions form an ideal)
that the derivatives and
generalised Lie derivatives of
an ancillary expression (expressed as $R^\fl(A,B)$)
is ancillary. This is what is needed to
consistently construct the brackets in the following Section.

The section property of $B_M$ implies that $\LL_{K^\fl}A=0$ when
$K^\fl$ is an ancillary expression (see eq. \eqref{LLXIsZero}).
This identity is also used in the
calculations for the identities of the brackets.

\section{The full \texorpdfstring{$L_\infty$}{BV}
  structure\label{FullBVSection}}

We will now display the full $L_\infty$ structure, including
ancillary ghosts.
The calculations for the $L_\infty$ brackets
performed in Section \ref{NonAncillaryBracketsSection} will be
revised in order to include
ancillary terms. 

\subsection{Some low brackets}

The $1$-bracket, which now acts on the ghosts $C$ at height $0$, and also
on ancillary ghosts $K$ at height $1$, is $d+\fl$:
\begin{align}
\dlb C+K\drb=dC+K^\fl+dK\;.
\end{align}
Since
$d^2=\fl^2=d\fl+\fl d=0$, the identity $\dlb\dlb C+K\drb\drb=0$ is
satisfied. The ancillary ghost at lowest level is automatically
annihilated by $d$.

The $2$-bracket identity was based on ``$d\LL_CC+\LL_CdC=0$'', which only
holds modulo ancillary terms. We need to modify the $2$-bracket to
\begin{align}
  \dlb C,C\drb&=\LL_CC+X_CC\;,\nn\\
  \dlb C,K\drb&=\tfrac12\LL_CK\;,\nn\\
  \dlb K,K\drb&=0\;.
\end{align}
Then,
\begin{align}
  &\dlb\dlb C,C\drb\drb+2\dlb C,\dlb C\drb\drb\nn\\
  &\qquad=\dlb\LL_CC+X_CC\drb+2\dlb C,dC\drb\nn\\
  &\qquad=d\LL_CC+X^\fl_CC+dX_CC+\LL_CdC+X_CdC=0\;.
\end{align}
thanks to eqs. \eqref{DefinitionOfX} and \eqref{dXPlusXd}, and
\begin{align}
  &\dlb\dlb C,K\drb\drb+2\cdot\tfrac12\dlb C,\dlb K\drb\drb
  +2\cdot\tfrac12\dlb K,\dlb C\drb\drb\nn\\
  &\qquad=\tfrac12\left(d\LL_CK+(\LL_CK)^\fl
  +\LL_CdK+\LL_CK^\fl+X_CK^\fl\right)=0\;.
\end{align}
The terms at height $1$ cancel using $X_CK^\fl=X^\fl_CK$, where the sign
follows from $\fl$ passing both a $d$ and an $\LL_C$.
Here, we have of course used $\LL_{K^\fl}=0$.
Note that the height $0$ identity involving one $K$ is trivial, while
the identity at height $1$ identity with one $K$ is equivalent to the
height $0$ identity with no $K$'s. These are both general features,
recurring in all bracket identities.
In addition $\dlb K,K^\fl\drb=\tfrac12\LL_{K^\fl}K=0$, implying that the
bracket with two $K$'s consistently can be set to $0$.

Consider the middle term in the $3$-bracket identity.
Including ancillary terms, we have
\begin{align}\label{3BracketMiddleTerm}
  &2\dlb C,\dlb C,C\drb\drb=2\dlb C,\LL_CC+X_CC\drb\nn\\
  &\qquad=\LL_C\LL_CC+X_C\LL_CC+\LL_{\LL_CC}C+X_{\LL_CC}C+\LL_CX_CC\nn\\
  &\qquad=\tfrac12(\LL_{\LL_CC}C+X_{\LL_CC}C)\;.
\end{align}
We know that $\dlb C,C,C\drb$ contains the non-ancillary term
$\tfrac13[C,\LL_CC]$. Calculating the contribution from this term to
$\dlb\dlb C,C,C\drb\drb+3\dlb C,C,\dlb C\drb\drb$ gives
\begin{align}
  &\tfrac13d[C,\LL_CC]+[C,\LL_CdC]+[dC,\LL_CC]\nn\\
  &\qquad=\tfrac13\left(-\tfrac32\LL_{\LL_CC}
  -[C,X_CC]^\fl-R^\fl(C,\LL_CC)\right)
  \end{align}
There is still no sign of something cancelling the second term in
eq. \eqref{3BracketMiddleTerm}, but the presence of lowered ancillary
terms implies that it is necessary to include the ancillary terms
$\tfrac13([C,X_CC]+R(C,\LL_CC))$ in the $3$-bracket. The term in
$\dlb\dlb C,C,C\drb\drb$ from the $\fl$ part of the $1$-bracket will
then cancel these. We still need to check the terms at height $1$.
The height $1$ contribution to
$\dlb\dlb C,C,C\drb\drb+3\dlb C,C,\dlb C\drb\drb$ from $\tfrac13[C,X_CC]$ is
\begin{align}
  \tfrac13\bigl(d[C,X_CC]+[C,X_CdC]+[dC,X_CC]\bigr)
  =\tfrac13\bigl(\LL_CX_CC+R(C,X^\fl_CC)\bigr)\;,
\end{align}
and from $\tfrac13R(C,\LL_CC)$, using eq. \eqref{RDerivationIdentity}:
\begin{align}
  &\tfrac13\bigl(dR(C,\LL_CC)+R(C,\LL_CdC)+R(dC,\LL_CC)\bigr)\nn\\
  &\qquad=\tfrac13\bigl(X_C\LL_CC-X_{\LL_CC}C-R(C,X^\fl_CC)\bigr)\;.
\end{align}
The complete height $1$ terms in the $3$-bracket identity become
\begin{align}
  (\tfrac12-\tfrac13)X_{\LL_CC}C+\tfrac13(\LL_CX_C+X_C\LL_C)C=0\;.
\end{align}

Checking the $3$-bracket identity with two $C$'s and one $K$ becomes
equivalent to the height $0$ identity for the bracket with three $C$'s
when
\begin{align}
\dlb C,C,K\drb=\tfrac19([C,\LL_CK]+[K,\LL_CC])\;.
\end{align}
There is also a height $0$ part of the $CCK$ identity, which is trivial
since $\fl$ generates no ancillary terms. Again, there is no need for
a bracket with $CKK$, since
\begin{align}
\dlb C,K,K^\fl\drb=\tfrac1{18}([K^\fl,\LL_CK]+[K,\LL_CK^\fl])=0\;.
\end{align}
These properties will be reflected at all orders, and we do not
necessarily mention them every time.

The $4$-bracket identity with four $C$'s reads
\begin{align}
  \dlb\dlb C,C,C,C\drb\drb+2\dlb C,\dlb C,C,C\drb\drb
  +3\dlb C,C,\dlb C,C\drb\drb+4\dlb C,C,C,\dlb C\drb\drb=0\;. 
\end{align}
We will now show that the vanishing of the $4$-bracket persists when
ancillary terms are taken into account.
The height $1$ terms in $2\dlb C,\dlb C,C,C\drb\drb$ are
\begin{align}
\tfrac13\bigl(X_C[C,\LL_C]+\LL_C[C,X_CC]+\LL_CR(C,\LL_CC)\bigr)\;,
\end{align}
and those in $3\dlb C,C,\dlb C,C\drb\drb$ become
\begin{align}
  &\tfrac13\bigl([C,X_C\LL_CC]+[C,\LL_CX_CC]+[C,X_{\LL_CC}C]
  +2[\LL_CC,X_CC]\nn\\
  &+\tfrac12R(C,\LL_{\LL_CC}C)+R(\LL_CC,\LL_CC)\bigr)
\end{align}
The terms cancel, using eqs. \eqref{RGenLieDerIdentity}
and \eqref{XGenLieDerIdentity}.

\subsection{Higher brackets}

The structure encountered so far can be extended to arbitrarily high
brackets. Knowing the height $0$ part of $\dlb C^n\drb=k_n(\ad
C)^{n-2}\LL_CC$ enables us to deduce the ancillary part. Namely,
keeping ancillary terms when applying eq.
\eqref{AlmostDerivationProperty} sequentially, calculating the first
and last terms in 
the $n$-bracket identity gives, apart from the second row of
eq. \eqref{FirstAndLastTerms},
\begin{align}
  d(\ad C)^{n-2}\LL_CC=\ldots
  -\Bigl((\ad C)^{n-2}X_cC
  +\sum\limits_{i=0}^{n-3}(\ad C)^iR_C(\ad C)^{n-i-3}\LL_CC\Bigr)^\fl\;.
\end{align}
This forces the $n$-bracket to take the form
\begin{align}
  \dlb C^n\drb=k_n\Bigl((\ad C)^{n-2}(\LL_CC+X_CC)
  +\sum\limits_{i=0}^{n-3}(\ad C)^iR_C(\ad C)^{n-i-3}\LL_CC\Bigr)\;.
\end{align}
It is then reasonable to assume that $\dlb C^{n-1},K\drb$ is obtained from
the symmetrisation of the height $0$ part of $\dlb C^n\drb$, \ie,
\begin{align}
  \dlb C^{n-1},K\drb=\frac{k_n}n\Bigl(
  (\ad C)^{n-2}\LL_CK+
  \sum\limits_{i=0}^{n-3}(\ad C)^i\ad K(\ad C)^{n-i-3}\LL_CC
  \Bigr)\;,
\end{align}
and that brackets with more than one $K$ vanish.

We will show that the set of non-vanishing brackets above is correct and complete.
The
height $0$ identity with only $C$'s is already satisfied, thanks to the
contribution from $\fl$ in $\dlb\dlb C^n\drb\drb$.
The height $1$ identity with one $K$ contains the same calculation.
The height $0$ identity with one $K$ is trivial, and just follows from
moving $\fl$'s in and out of commutators and through derivatives and
generalised Lie derivatives. The vanishing of the brackets with more
than one $K$ is consistent with the vanishing of
$\dlb C^{n-2},K^\fl,K\drb$. Lowering this bracket gives
$\dlb C^{n-2},K^\fl,K^\fl\drb$ which vanishes by statistics, since
$K^\fl$ is fermionic.

The only remaining non-trivial check is the height $1$ part of the
identity with only $C$'s.
This is a lengthy calculation that relies on all identities exposed in
Section \ref{OperatorsSection}. We will go through the details by
collecting the different types of terms generated, one by one.

A first result of the calculation is that all terms containing more
than one ancillary expression $X$ or $R$ cancel. This
important consistency condition relies on the
precise combination of terms in the $n$-bracket,
but not on the relation between the
coefficients $k_n$. It could have been used as an alternative means to
obtain possible brackets.

We then focus on the terms containing $X$.
In addition to its appearance in the brackets, $X$ arises when a
derivative or a generalised Lie derivative is taken through an $R$,
according to eqs. \eqref{RDerivationIdentity} and
\eqref{RGenLieDerIdentity}. It
turns out that all terms where $X_C$ appears in an ``inner'' position
in terms of the type 
$(\ad C)^iX_C(\ad C)^{n-i-3}\LL_CC$, with $n-i>3$, cancel. This
again does not depend on the coefficients $k_n$.
Collecting terms $(\ad C)^{n-3}\LL_CX_CC$ and
$(\ad C)^{n-3}X_C\LL_CC$, the part $\dlb\dlb C^n\drb\drb+n\dlb C^{n-1},\dlb C\drb\drb$
gives a contribution
\begin{align}
  k_n(n-2)(\ad C)^{n-3}\LL_CX_CC
\end{align}
from the $X$ term in the bracket, and
\begin{align}
  k_n\bigl((n-2)(\ad C)^{n-3}X_C\LL_CC-(\ad C)^{n-3}X_{\LL_CC}C\bigr)
\end{align}
from the
$R$ term, together giving
\begin{align}
  -\frac n2k_n(\ad C)^{n-3}X_{\LL_CC}C\;.
\end{align}
A middle term in the identity, $\dlb C^i,\dlb C^{n-i}\drb\drb$
contains
\begin{align}
  -\tfrac12k_{i+1}k_{n-i}(\ad C)^{n-3}X_{\LL_CC}C\;.
\end{align}
The total contribution cancels, thanks to the relation
\eqref{FirstCoeffRelation} between the coefficients.

The remaining terms with $X$ are of the types
$(\ad C)^j\ad\LL_CC(\ad C)^{n-4-j}X_CC$ and 
$(\ad C)^j\ad X_CC(\ad C)^{n-4-j}\LL_CC$ and similar.
The first and last term in the identity gives a contribution
\begin{align}
-k_n\sum\limits_{j=0}^{n-4}(j+1)(\ad C)^j\ad\LL_CC(\ad C)^{n-4-j}X_CC
\end{align}
from the $X$ term in the $n$-bracket, and
\begin{align}
-k_n\sum\limits_{j=0}^{n-4}(j+1)(\ad C)^j\ad X_CC(\ad C)^{n-4-j}\LL_CC
\end{align}
from the $R$ term.
A middle term $\dlb C^i,\dlb C^{n-i}\drb\drb$ gives
\begin{align}
  k_{i+1}k_{n-i}\Bigl(
&-\sum\limits_{j=0}^{n-i-3}(\ad C)^{i+j-1}\ad \LL_CC(\ad C)^{n-i-j-3}X_CC
  \Bigr.\nn\\
&-\sum\limits_{j=0}^{n-i-3}(\ad C)^{i+j-1}\ad X_CC(\ad C)^{n-i-j-3}\LL_CC\nn\\
&+\sum\limits_{j=0}^{i-2}(\ad C)^j\ad((\ad C)^{n-i-2}\LL_CC)(\ad C)^{i-j-2}X_CC\nn\\
\Bigl.
&+\sum\limits_{j=0}^{i-2}(\ad C)^j\ad((\ad C)^{n-i-2}X_CC)(\ad C)^{i-j-2}\LL_CC
\Bigr)\;.
\end{align}
Note the symmetry between $X_C$ and $\LL_C$ in all contributions.
We can now represent a term
$(\ad C)^{n-4-j-k}[(\ad C)^j\LL_CC,(\ad C)^kX_CC]$ by a monomial
$s^jt^k$, exactly as in Section
\ref{HigherNonAncillaryBracketsSection}. Since we have the symmetry
under $s\leftrightarrow t$, the same rules apply as in that calculation.
Indeed, precisely the same polynomials are generated as in eqs.
\eqref{KnPolynomial} and \eqref{MiddlePolynomials}. The terms cancel.

Finally, there are terms of various structure
with one $R$ and two $\LL$'s. One such
structure is $(\ad C)^jR_C(\ad C)^{n-j-4}\LL_{\LL_CC}C$. For each value
of $j$, the total coefficient of the term cancels thanks to
$nk_n+\sum_ik_{i+1}k_{n-i}=0$.
Of the remaining terms, many have $C$ as one of the two arguments of
$R$, but some do not. In order to deal with the latter, one needs the
cyclic identity \eqref{CyclicIdentity}.
Let
\begin{align}
F_j=(\ad C)^j\LL_CC \qquad \text{and} \qquad
S_{n,j,k}=(\ad C)^{n-j-k-4}R(F_j,F_k)\,.
\end{align}
Taking the arguments in the cyclic identity as $C$, $F_j$ and $F_k$
turns it into
\begin{align}
  S_{n,j,k}-S_{n,j+1,k}-S_{n,j,k+1}
  =-(\ad C)^{n-j-k-5}\bigl(R_C[F_j,F_k]-2[F_{(j},R_CF_{k)}]\bigr)\;.
  \label{WeakSTEquivalence}
\end{align}
We need to verify that terms containing $S_{n,j,k}$, \ie, not having $C$
as one of the arguments of $R$, combine into the first three terms of
this equations, and thus can be turned into expressions with $R_C$.
Note that this relation is analogous to
eq. \eqref{CyclicIdentity} for $Z_{n,j,k}$ in Section
\ref{HigherNonAncillaryBracketsSection}, but with a remainder term.
We now collect such terms. They are
\begin{align}
  -k_n\sum\limits_{j=0}^{n-4}(n-3-j)S_{n,j,0}
  +\sum\limits_{i=2}^{n-2}k_{i+1}k_{n-i}\Bigl(
\sum\limits_{j=0}^{i-2}S_{n,j,n-i-2}-\sum\limits_{j=0}^{n-i-3}S_{n,j,0}
\Bigr)\;.
\label{STermsCancellation}
\end{align}
This is the combination encountered earlier
(eqs. \eqref{KnPolynomial} and \eqref{MiddlePolynomials}),
which means that these
terms can be converted to terms with $R_C$. However, since the
``$s+t-1\approx0$''
relation in the form \eqref{WeakSTEquivalence}
now holds only modulo $R_C$ terms, we need to add the corresponding
$R_C$ terms to the ones already present. 

Let us now proceed to the last remaining terms. They are of two types:
\begin{align}
U_{n,r,j,k}&=(\ad C)^rR_C(\ad C)^{n-r-j-k-5}[(\ad C)^j\LL_CC,(\ad
  C)^k\LL_CC]\;,\nn\\
V_{n,r,j,k}&=(\ad C)^{n-r-j-k-5}[(\ad C)^j\LL_CC,(\ad C)^kR_C(\ad C)^r\LL_CC]\;.
\end{align}
If the $j$ and $k$ indices in both expressions are translated into
monomials $s^jt^k$ as before, both expressions should be calculated
modulo $s+t-1\approx0$ as before. In $U$, symmetry under
$s\leftrightarrow t$ can be used, but not in $V$. Both types of terms
need to cancel for all values of $r$, since there is no identity that
allows us to take $\ad C$ past $R_C$.

The terms of type $U_{n,r,j,k}$ obtained directly from
$\dlb\dlb C^n\drb\drb+n\dlb C^{n-1},\dlb C\drb\drb$ are
\begin{align}
-k_n\sum\limits_{r=0}^{n-5}\sum\limits_{k=0}^{n-r-5}(n-k-3)U_{n,r,0,k}\;,
\end{align}
and those from $\dlb C^i,\dlb C^{n-i}\drb\drb$ are
\begin{align}
  k_{i+1}k_{n-i}\Bigl(
  \sum\limits_{r=0}^{i-3}\sum\limits_{k=0}^{i-r-3}U_{n,r,n-i-2,k}
  -\sum\limits_{r=0}^{i-2}\sum\limits_{k=0}^{n-i-3}U_{n,r,0,k}
  -\sum\limits_{r=i-1}^{n-5}\sum\limits_{k=0}^{n-r-5}U_{n,r,0,k}\;
  \Bigr)\,.
\end{align}
To these contributions must be added the remainder term corresponding
to the first term on the right hand side of
eq. \eqref{WeakSTEquivalence}, with the appropriate coefficients from
eq. \eqref{STermsCancellation}.
Let $U_{n,r,j,k}$ correspond to the monomial $s^jt^ku^r$. According
to eq. \eqref{WeakSTEquivalence}, the
remainder terms then become 
\begin{align}
  u^{n-5}\frac{\phi_n(\frac su,\frac tu)}{\frac su+\frac tu-1}
  \approx\frac{u^{n-4}}{1-u}\phi_n\bigl(\frac su,\frac tu\bigr)\;,
\end{align}
where $\phi(s,t,x)=\sum_{n=2}^\infty\phi_n(s,t)x^n$, and
where $s+t-1\approx0$ has been used in the last step.
The total contribution to the $n$-bracket identity then is
\begin{align}
&\tfrac12\Bigl[
  k_n\Bigl(-\sum\limits_{r=0}^{n-5}\sum\limits_{k=0}^{n-r-5}(n-k-3)u^rt^k
  -\frac{u^{n-4}}{1-u}\sum\limits_{k=0}^{n-4}(n-k-3)
  \bigl(\frac tu\bigr)^k\Bigr)
  \Bigr.\nn\\
&\qquad  +\sum\limits_{i=1}^{n-2}k_{i+1}k_{n-i}\Bigl(
  \sum\limits_{r=0}^{i-3}\sum\limits_{k=0}^{i-r-3}s^{n-i-2}t^ku^r
  -\sum\limits_{r=0}^{i-2}\sum\limits_{k=0}^{n-i-3}t^ku^r
  -\sum\limits_{r=i-1}^{n-5}\sum\limits_{k=0}^{n-r-5}t^ku^r\Bigr.\nn\\
&\Bigl.\Bigl.\qquad  +\frac{u^{n-4}}{1-u}
  \sum_{k=0}^{i-2}\bigl(\frac su\bigr)^{n-i-2}\bigl(\frac tu\bigr)^k
  -\frac{u^{n-4}}{1-u}\sum\limits_{k=0}^{n-i-3}\bigl(\frac tu\bigr)^k
  \Bigl)\Bigr]+(s\leftrightarrow t)\nn\\
  &\quad=\tfrac12\Bigl[
    -k_n\frac{n-3-(n-2)t+t^{n-2}}{(1-t)^2(1-u)}\Bigr.\nn\\
      &\Bigl.\qquad+\sum\limits_{i=1}^{n-2}k_{i+1}k_{n-i}
      \frac{(1-t^{i-1})s^{n-i-2}-(1-t^{n-i-2})}{(1-t)(1-u)}
      \Bigr]+(s\leftrightarrow t)\nn\\
 &\quad=\frac{\phi_n(s,t)}{1-u}\;.
\end{align}
The $U_{n,r,s,t}$ terms thus cancel for all values of $r$.

The terms of type $V_{n,r,s,t}$ obtained directly from
$\dlb\dlb C^n\drb\drb+n\dlb C^{n-1},\dlb C\drb\drb$ are
\begin{align}
  -k_n\sum\limits_{r=0}^{n-5}\sum\limits_{k=0}^{n-r-5}
  (n-r-k-4)V_{n,r,0,k}
\end{align}
and the ones from $\dlb C^i,\dlb C^{n-i}\drb\drb$ are
\begin{align}
  k_{i+1}k_{n-i}\Bigl(
  &-\sum\limits_{r=0}^{n-i-4}\sum\limits_{k=0}^{n-i-r-4}V_{n,r,0,k}
  +\sum\limits_{j=0}^{i-2}\sum\limits_{k=0}^{n-i-3}V_{n,n-i-k-3,j,k}\nn\\
  &+\sum\limits_{r=0}^{i-3}\sum\limits_{k=0}^{i-r-3}V_{n,r,n-i-2,k}
  \Bigr)\;.
\end{align}
In addition, there is a remainder term from the second term on the
right hand side of 
eq. \eqref{WeakSTEquivalence}. If $V_{n,r,j,k}$ is represented by
$s^jt^ku^r$, the remainder term becomes
\begin{align}
-2\frac{\phi_n(s,u)}{s+u-1}\;.
\end{align}
The total contribution of terms of type $V$ to the $n$-bracket is then
represented by the function $v_n(s,t,u)$:
\begin{align}
  v_n(s,t,u)&=-k_n\sum\limits_{r=0}^{n-5}\sum\limits_{k=0}^{n-r-5}
  (n-r-k-4)t^ku^r\nn\\
  &\quad\,+\sum\limits_{i=1}^{n-2}k_{i+1}k_{n-i}\Bigl(
  -\sum\limits_{r=0}^{n-i-4}\sum\limits_{k=0}^{n-i-r-4}t^ku^r
  +\sum\limits_{j=0}^{i-2}\sum\limits_{k=0}^{n-i-3}s^jt^ku^{n-i-k-3}\nn\\
  &\qquad\qquad\qquad\qquad
         +\sum\limits_{r=0}^{i-3}\sum\limits_{k=0}^{i-r-3}s^{n-i-2}t^ku^r
  \Bigr)\\
  &\quad\,+\frac1{s+u-1}\Bigl[k_n\sum\limits_{\ell=0}^{n-4}(n-\ell-3)(s^\ell+u^\ell)
    \Bigr.\nn\\
    &\Bigl.\qquad\qquad\qquad+\sum\limits_{i=1}^{n-2}k_{i+1}k_{n-i}\Bigl(
    \sum\limits_{\ell=0}^{n-i-3}(s^\ell+u^\ell)
    -\sum\limits_{\ell=0}^{i-2}(s^{n-i-2}u^\ell+s^\ell
    u^{n-i-2})\Bigr)\Bigr]\;.
  \nn
\end{align}
Performing the sums, except the ones over $i$, and replacing $s$ by
$1-t$, this function turns into
\begin{align}
v_n(1-t,t,u)=2\frac{\phi_n(1-t,t)}{t-u}\;.
\end{align}
Therefore, these terms cancel. Note that the symmetrisation
$s\leftrightarrow t$ in $\phi_n$ is automatic, and not imposed by hand.
This concludes the proof that all the identities are satisfied.

The series $\sum_{n=2}^\infty k_n(\ad C)^{n-2}$
appearing in the variation of the ghosts, the sum of all brackets, can
be written in the concise form $g(\ad C)$, where
$g(x)=\frac2{1-e^{-2x}}-\frac1x$. Likewise, the sum
$\sum_{n=2}^\infty \frac{k_n}n(\ad C)^{n-2}$ becomes $h(\ad C)$, where
\begin{align}
  h(x)=\frac1{x^2}\int\limits_0^xdy\,yg(y)
  =1-\frac1x+\log(1-e^{-2x})
  -\frac1{x^2}\bigl(\hbox{Li}_2(e^{-2x})-\frac{\pi^2}{12}\bigr)
  \;.
\end{align}
The terms in the brackets containing sums of type $\sum_{i=0}^{n-3}(\ad C)^i{\scr
  O}(\ad C)^{n-i-3}$ 
can be formally rewritten, \eg,
\begin{align}
  \sum\limits_{n=2}^\infty k_n\sum\limits_{i=0}^{n-3}(\ad C)^i{\scr O}(\ad
  C)^{n-i-3}
  &=\sum\limits_{n=2}^\infty k_n
    \frac{(\ad C)_L^{n-2}-(\ad C)_R^{n-2}}{(\ad C)_L-(\ad C)_R}
         {\scr O}\nn\\
  &=\frac{g((\ad C)_L)-g((\ad C)_R)}{(\ad C)_L-(\ad C)_R}{\scr O}\;,
\end{align}
where subscripts $L,R$ stands for action to the left or to the right
of the succeeding operator (${\scr O}$).
Then, the full ghost variation takes the functional form
\begin{align}
  &(S,C+K)=(d+\fl)(C+K)+g(\ad C)(\LL_C+X_C)C+h(\ad C)\LL_CK\\
  &\quad+\Bigl[
    \frac{g((\ad C)_L)-g((\ad C)_R)}{(\ad C)_L-(\ad
      C)_R}R_C\Bigr]\LL_CC
  +\Bigl[
    \frac{h((\ad C)_L)-h((\ad C)_R)}{(\ad C)_L-(\ad C)_R}\ad K\Bigr]\LL_CC
  \;.\nn
\end{align}

\section{Examples\label{ExamplesSection}}

The criterion that no ancillary transformations appear in the
commutator of two generalised diffeomorphisms is quite restrictive. It
was shown in ref. \cite{Cederwall:2017fjm} that this happens if and
only if $\fg_r$ is finite-dimensional and the derivative module is
$R(\lambda)$ where $\lambda$ is a fundamental weight dual to
a simple root with Coxeter label $1$. The complete list is
\begin{itemize}
\item[({\it i})] $\fg_r=A_r$, $\lambda=\Lambda_p$, $p=1,\ldots,r$
($p$-form representations);
\item[({\it ii})] $\fg_r=B_r$, $\lambda=\Lambda_1$ (the vector representation);
\item[({\it iii})] $\fg_r=C_r$, $\lambda=\Lambda_r$ (the 
symplectic-traceless $r$-form representation);
\item[({\it iv})] $\fg_r=D_r$,
$\lambda=\Lambda_1,\Lambda_{r-1},\Lambda_r$ (the vector and spinor
representations);
\item[({\it v})] $\fg_r=E_6$, $\lambda=\Lambda_1,\Lambda_5$ (the
fundamental representations);
\item[({\it vi})] $\fg_r=E_7$, $\lambda=\Lambda_1$ (the fundamental
representation).
\end{itemize}
If $\fg_{r+1}$ has a $5$-grading or higher with respect to the
subalgebra $\fg_r$ (in particular, if it is
infinite-dimensional), $\tR_2$ will be non-empty (see Table
\ref{GeneralTable}), and there will be ancillary ghosts starting from
level $1$ (ghost number $2$).

Ordinary diffeomorphisms provide a simple and quite
degenerate example, where $\fg_r=A_r$ and $\lambda=\Lambda_1$.
In this case, both $R_2$ and
$\tR_2$ are empty, so 
both $\fg_{r+1}$ and ${\scr B}(\fg_r)$ are 3-gradings. Still, the
example provides the core of all other examples. The algebra of vector
fields in $r+1$ dimensions is constructed using the
structure constants of 
\begin{align}
{\scr B}(A_{r+1})\approx A(r+1|0)\approx\mathfrak{sl}(r+2|1)\,.
\end{align}
There is of course neither any reducibility nor any ancillary ghosts,
and the only ghosts are the ones in the vector representation
${\bf v}$ in 
$R_{(1,0)}$. The double grading of the superalgebra is given in Table
\ref{DiffeomorphismsTable}. 

\begin{table}
\begin{align*}
  \xymatrix{
    \ar@{-}[]+<1.1cm,1em>;[ddd]+<1.1cm,-1em>
    \ar@{-}[]+<-0.8cm,-1em>;[rrr]+<0.6cm,-1em>
&p=-1 &p=0 &p=1\\
q=1&&{\bf 1}   &    {\bf v}  \\
q=0&\overline {\bf v} &{\bf 1}\oplus{\bf adj}\oplus{\bf 1}  & {\bf v}  \\
q=-1 &\overline {\bf v} & {\bf 1} & 
  }
\end{align*}
\caption{\it The decomposition of $A(r+1|0)\approx\mathfrak{sl}(r+2|1)$
  in $A(r)\approx\mathfrak{sl}(r+1)$ modules.}
\label{DiffeomorphismsTable}
\end{table}

The double diffeomorphisms, obtained from $\fg_r=D_r$,
have a singlet reducibility, and no
ancillary transformations. The $L_\infty$ structure (truncating to an
$L_3$ algebra) was examined in
ref. \cite{Hohm:2017pnh}. The Borcherds superalgebra is finite-dimensional,
\begin{align}
{\scr B}(D_{r+1})\approx D(r+1|0)\approx\mathfrak{osp}(r+1,r+1|2)\,.
\end{align}
The double grading of this superalgebra is given in Table
\ref{DoubleDiffeomorphismsTable}. The only ghosts are the (double) vector 
in $R_{(1,0)}$ and the singlet in $R_{(2,0)}$.

The extended geometry based on $\fg_r=B_r$ follows an analogous
pattern, and is also described by Table
\ref{DoubleDiffeomorphismsTable}, but with the doubly extended
algebra
$B(r+1,0)\approx\mathfrak{osp}(r+1,r+2|2)$ being decomposed into
modules of $B(r)\approx\mathfrak{so}(r,r+1)$.

Together with the ordinary diffeomorphisms, these are the only cases
with finite reducibility and without ancillary transformations at
ghost number $1$. In order for the reducibility to be finite, it is
necessary that ${\scr B}(\fg_r)$ is finite-dimensional. The remaining
finite-dimensional superalgebras in the classification by Kac
\cite{Kac77B} are not represented by Dynkin diagrams where
the grey node connects to a node
with Coxeter label $1$. Therefore, even if there are other examples
with finite-dimensional ${\scr B}(\fg_r)$, they all have ancillary
transformations appearing in the commutator of two generalised Lie
derivatives. Such examples may be interesting to investigate in the
context of the tensor hierarchy algebra (see the discussion in Section
\ref{ConclusionsSection}).

\begin{table}
\begin{align*}
  \xymatrix{
    \ar@{-}[]+<1.1cm,1em>;[ddd]+<1.1cm,-1em>
    \ar@{-}[]+<-0.8cm,-1em>;[rrrrr]+<0.6cm,-1em>
&p=-2&p=-1 &p=0 &p=1&p=2\\
q=1&&&{\bf 1}   &    {\bf v} &{\bf 1} \\
q=0&{\bf1}&{\bf v} &{\bf 1}\oplus{\bf adj}\oplus{\bf 1}  & {\bf v}&{\bf1}  \\
q=-1 &{\bf 1}& {\bf v} & {\bf 1} & &
  }
\end{align*}
\caption{\it The decomposition of $D(r+1|0)\approx{\mathfrak{
    osp}}(r+1,r+1|2)$ in $D(r)\approx\mathfrak{so}(r,r)$ modules.}
\label{DoubleDiffeomorphismsTable}
\end{table}

We now consider the cases $\fg_r=E_r$ for $r\leq7$.
The level decompositions of the Borcherds superalgebras are described in
ref. \cite{Cederwall:2015oua}. There are always ancillary ghosts,
starting at level $8-r$ (ghost number $9-r$).
In Table \ref{ExceptionalDiffTable}, we give the double grading in the
example $\fg_r=E_{5(5)}\approx \mathfrak{so}(5,5)$. Modules $\tR_p$ are
present for $p\geq4$, signalling an infinite tower of ancillary ghost
from ghost number $4$. Table \ref{ExceptionalDiffTableE7} gives the
corresponding decomposition for $\fg_r=E_{7(7)}$. This is as far as
the construction of the present paper applies. Note that for $\fg_r=E_{7(7)}$ already
$\tR_2={\bf1}$, which leads to ancillary ghosts in the ${\bf56}$ at
$(p,q)=(1,1)$. 
\begin{table}
  \begin{align*}
  \xymatrix@=.4cm{
    \ar@{-}[]+<1cm,1em>;[dddd]+<1cm,-1em>
    \ar@{-}[]+<-0.8cm,-1em>;[rrrrrrr]+<1.0cm,-1em>
&p=-1 &p=0 &p=1&p=2&p=3&p=4&p=5\\
q=2&&&&&&{\bf1}&{\bf16}\\
q=1&&{\bf 1}   &    {\bf16} &{\bf10}&\overline{\bf16}
           &{\bf45}\oplus{\bf1}&\overline{\bf144}\oplus{\bf16}\\
           q=0&\overline{\bf16}&{\bf 1}\oplus{\bf45}\oplus{\bf 1}
           & {\bf16}&{\bf10}&\overline{\bf16}&{\bf45}&\overline{\bf144}  \\
q=-1 & \overline{\bf16} & {\bf 1} 
  }
\end{align*}
\caption{\it Part of the decomposition of ${\scr B}(E_{6(6)})$ in
  $E_{5(5)}\approx\mathfrak{so}(5,5)$ modules. Note the appearance of
modules $\tR_p$ for $p\geq4$.}
\label{ExceptionalDiffTable}
\end{table}
\begin{table}
  \begin{align*}
  \xymatrix@=.3cm{
    \ar@{-}[]+<.9cm,1em>;[ddddd]+<.9cm,-1em>
    \ar@{-}[]+<-0.8cm,-1em>;[rrrrr]+<2.6cm,-1em>
&p=0 &p=1&p=2&p=3&p=4\\
q=3&&&&&{\bf1}\\
q=2&&&{\bf1}&{\bf56}&{\bf1539}\oplus{\bf133}\oplus2\cdot{\bf1}\\
q=1&{\bf1}&{\bf56}&{\bf133}\oplus{\bf1}&{\bf912}\oplus{\bf56}
        &{\bf8645}\oplus2\cdot{\bf133}\oplus{\bf1539}\oplus{\bf1}\\
q=0&{\bf1}\oplus{\bf133}\oplus{\bf 1}
           & {\bf56}&{\bf133}&{\bf912}&{\bf8645}\oplus{\bf133}& \\
q=-1 & {\bf1} 
  }
\end{align*}
\caption{\it Part of the decomposition of ${\scr B}(E_{8(8)})$ in
  $E_{7(7)}$ modules.}
\label{ExceptionalDiffTableE7}
\end{table}
In Table \ref{ExceptionalDectionDiffTable}, we have divided the
modules $R_p$ for the $E_{5(5)}$ example of Table
\ref{ExceptionalDiffTable}
into $A_4$ modules with respect to a choice of section.
Below the solid dividing line are the usual sequences of ghosts for
diffeomorphisms and $2$-form and $5$-form gauge transformations.
Above the line are sequences that contain tensor products of forms
with some other modules, \ie, mixed tensors.
All modules above the line are effectively
cancelled by the ancillary ghosts. They are however needed to build
modules of $\fg_r$.
In the example, there is nothing below the line for $p\geq7$, which
means that the $\fl$ operation from ancillary to non-ancillary ghosts
at these levels becomes bijective.

Reducibility is of course not an absolute concept; it can depend on the
amount of covariance maintained. If a section is chosen, the
reducibility can be made finite by throwing away all ghosts above the
dividing line.
One then arrives at the situation in ref. \cite{Baraglia:2011dg}.
If full covariance is maintained, reducibility is infinite.  
Since the modules above the line come in tensor
products of some modules with full sets
of forms of alternating statistics, they do not contribute to the
counting of the degrees of freedom. This shows why the counting of
refs. \cite{Berman:2012vc,Cederwall:2015oua},
using only the non-ancillary ghosts, gives
the correct counting of the number of independent gauge parameters.

This picture of the reduction of the modules $R_p$ in a grading with
respect to the choice of section also makes the characterisation of
ancillary ghosts clear. They are elements in $R_p$ above a certain
degree (for which the degree of the derivative is $0$). The dotted
line in the table indicates degree $0$. If we let ${\scr A}$ be the
subalgebra of ancillary elements above the solid line, it is clear
that ${\scr A}$ forms an ideal in ${\scr B}_+(\fg_r)$ (which was also
shown on general grounds in Section \ref{AncillaryGhostsSection}).
The grading coincides with the
grading used in ref. \cite{Cederwall:2017fjm} to show that the
commutator of two ancillary transformations again is ancillary.

\begin{landscape}
\begin{table}
  \begin{align*}
  \xymatrix@=.4cm@C=0.8cm{
    \ar@{-}[]+<2em,1em>;[ddddddd]+<2em,-1em>
    \ar@{-}[]+<-0.5cm,-1em>;[rrrrrr]+<2.4cm,-1em>
    \ar@{-}[]+<-0.8cm,-4.5cm>;[rrrrrr]+<2.4cm,-4.5cm>
    &p=1 &p=2 &p=3&p=4&p=5&p=6\\
v=6&&&&&&(\overline{\bf15}\oplus{\bf40})\otimes\Lambda_5\\
    v=5&&&&&{\bf 24}\otimes\Lambda_5&{\bf 24}\otimes\Lambda_4
    \oplus({\bf5}\oplus\overline{\bf45})\otimes\Lambda_5\\
    v=4&&&&{\bf10}\otimes\Lambda_5&{\bf10}\otimes\Lambda_4\oplus \overline{\bf15}\otimes\Lambda_5&{\bf10}\otimes\Lambda_3
    \oplus \overline{\bf15}\otimes\Lambda_4 \oplus   \overline{\bf5}\otimes\Lambda_5\\ 
v=3&&&\overline{\bf5}\otimes\Lambda_5&\overline{\bf5}\otimes\Lambda_4&\overline{\bf5}\otimes\Lambda_3&\overline{\bf5}\otimes\Lambda_2\\
v=2&\Lambda_5&\Lambda_4&\Lambda_3&\Lambda_2&\Lambda_1&\Lambda_0\\
v=1&\Lambda_2&\Lambda_1&\Lambda_0\\
v=0&\Lambda_4
}
\end{align*}
\caption{\it Part of the decomposition of $R_p$ for the $E_{5(5)}$
  exceptional geometry with respect to a section $\mathfrak{sl}(5)$.
  The derivative acts horizontally to the left and
  $\Lambda_k$ denote the $k$-form modules of $\mathfrak{sl}(5)$, such that $\Lambda_1,\Lambda_2,\ldots=\overline{\bf 5},\overline{\bf 10},\ldots$
and $\Lambda_0=\Lambda_5={\bf 1}$. The degree $v$ is such that the relative weights in the extension to $\mathfrak{gl}(5)$ are given by $3v+4p$. The $\Lambda_4$
in the lower left corner is the vector module corresponding to the ordinary coordinates with this choice of section.}
\label{ExceptionalDectionDiffTable}
\end{table}
\end{landscape}

As an aside, the regularised dimension, twisted with fermion
number, of ${\scr B}_+(\fg_{r+1})$ can readily be calculated using the
property that all modules at $p\neq0$ come in doublets under the
superalgebra generated by $e$ and $f$, without need of any further
regularisation (\eg\ through analytic continuation).
Using the cancellation of these
doublets, inspection of Table \ref{GeneralTable} gives at hand that
the ``super-dimension'' (where fermionic generators count with a minus sign)
\begin{align}
-\sdim({\scr B}_+(\fg_{r+1}))
  &=1+\dim(R_1)+\dim(\tR_2)+\dim(\ttR_3)+\ldots\nn\\
  &=1+\dim(\fg_{r+1,+})\;,
\end{align}
where $\fg_{r+1,+}$ is the positive level part of the grading of
$\fg_{r+1}$ with respect to $\fg_r$. This immediately reproduces the
counting of the effective number of gauge transformations in
ref. \cite{Berman:2012vc}. In the example ${\scr B}(E_6)$
above, we get $1+16=17$,
which is the correct counting of gauge parameters for diffeomorphisms,
$2$- and $5$-form gauge transformations in $6$ dimensions.

\section{Conclusions\label{ConclusionsSection}}

We have provided a complete set of bracket giving an $L_\infty$
algebra for generalised diffeomorphisms in extended geometry,
including double geometry and exceptional geometry as special cases.
The construction depends crucially on the use of the underlying Borcherds
superalgebra ${\scr B}({\mathfrak g}_{r+1})$,
which is a double extension of the structure algebra
${\mathfrak g}_r$. This superalgebra is needed in order to form the
generalised diffeomorphisms, and has a natural interpretation in terms
of the section constraint. It also provides a clear criterion for the
appearance of ancillary ghosts.

The full list of non-vanishing brackets is:
\begin{align}
\dlb C\drb&=dC\;,\nn\\
\dlb K\drb&=dK+K^\fl\;,\nn\\
\dlb C^n\drb&=k_n\Bigl((\ad C)^{n-2}(\LL_CC+X_CC)
+\sum\limits_{i=0}^{n-3}(\ad C)^iR_C(\ad C)^{n-i-3}\LL_CC\Bigr)\;\\
\dlb C^{n-1},K\drb&=\frac{k_n}n\Bigl(
  (\ad C)^{n-2}\LL_CK+
  \sum\limits_{i=0}^{n-3}(\ad C)^i\ad K(\ad C)^{n-i-3}\LL_CC
  \Bigr)\;,\nn
\end{align}
where the coefficients have the universal model-independent expression
in terms of Bernoulli numbers
\begin{align}
  k_{n+1}=\frac{2^nB^+_n}{n!}\;,\quad n\geq1\;.
\end{align}
All non-vanishing 
brackets except the $1$-bracket contain at least one level $1$ ghost $c$.
No brackets contain more than one ancillary ghost.

The violation of covariance of the derivative,
that modifies already the $2$-bracket,
has a universal form, encoded in $X_C$ in eq. \eqref{ExplicitFormOfX}. 
It is not unlikely that this makes it possible to covariantise the
whole structure, as in ref. \cite{Cederwall:2015ica}.
However, we think that it is appropriate to let the algebraic
structures guide us concerning such issues.

The characterisation of ancillary ghosts is an interesting issue, that
may deserve further attention. Even if the construction in Section
\ref{AncillaryGhostsSection} makes the appearance of ancillary ghosts clear
(from the existence of modules $\tR_p$) it is indirect and does not
contain an independent characterisation of the ancillary ghosts, in
terms of a constraint. This property is shared with the construction
of ancillary transformations in ref. \cite{Bossard:2017aae}.
The characterisation in Section \ref{ExamplesSection}
in terms of the grading induced by a choice of
section is a direct one, in this sense, but has the drawback that it
lacks full covariance. In addition, there may be more than one
possible choice of section. This issue may become more important when
considering situations with ancillary ghosts at ghost number $1$ (see
below). Then, with the exception of some simpler cases with
finite-dimensional $\fg_r$,
ancillary transformations are not expected to commute.

We have explicitly excluded from our analysis cases where ancillary
transformations appear already at ghost number $1$
\cite{Hohm:2013jma,Hohm:2014fxa,Cederwall:2015ica,Bossard:2017aae}. The
canonical example is exceptional geometry with structure group
$E_{8(8)}$. If we should trust and extrapolate the results of the 
present paper, this would correspond to the presence of a module
$\tR_1$. However, there is never such a module in the Borcherds
superalgebra.
If we instead turn to the tensor hierarchy algebra
\cite{Palmkvist:2013vya,Carbone:2018njd,Bossard:2017wxl}
we find that a module $\tR_1$ indeed appears in cases when ancillary
transformations are present in the commutator of two generalised
diffeomorphisms. 

As an example, Table \ref{E8Table} contains a part of the double grading of the
tensor hierarchy algebra $W(E_9)$ (following the notation of ref. \cite{Carbone:2018njd}),
which we believe should be used in the
construction of an $L_\infty$ algebra for $E_8$ generalised
diffeomorphisms.
The $E_8$ modules that are not present in the ${\scr B}(E_9)$
superalgebra are marked in blue colour. The singlet at $(p,q)=(1,1)$ is
the extra element
appearing at level $0$ in $W(E_9)$ 
that can be identified with the Virasoro generator $L_1$
(as can be seen in the decomposition under $\mathfrak{gl}(9)$ \cite{Bossard:2017wxl}).
The elements at $q-p=1$ come from the ``big''
module at level $-1$ in $W(E_9)$ (the embedding tensor or big torsion
module). For an affine $\fg_{r+1}$ this is a shifted fundamental
highest weight module, with its highest weight at $(p,q)=(1,2)$,
appearing in $W(\fg_{r+1})$ in
addition to the unshifted one with highest weight at $(p,q)=(0,1)$
appearing also in the Borcherds superalgebra ${\scr B}(\fg_{r+1})$.
In the $E_8$ example, it
contains the ${\bf248}$ at $(p,q)=(0,1)$ which will accommodate
parameters of the ancillary transformations.
In situations when ancillary transformations are absent at ghost
number $1$ (the subject of the present paper), using $W(\fg_{r+1})$ is
equivalent to using ${\scr B}(\fg_{r+1})$, so all results derived here
will remain unchanged.

We take this as a very strong sign that the tensor
hierarchy algebra is the correct underlying algebra, and
hope that a generalisation of the present approach to the use of an
underlying tensor
hierarchy algebra will shed new light on the properties of generalised
diffeomorphisms in situations where ancillary transformations are present.

\begin{table}
  \begin{align*}
  \xymatrix@=.4cm{
    \ar@{-}[]+<1.1cm,1em>;[dddd]+<1.1cm,-1em>
    \ar@{-}[]+<-0.8cm,-1em>;[rrrr]+<1.4cm,-1em>
&p=-1 &p=0 &p=1&p=2\\
q=2&&&\Blue{\bf1}&{\bf248}\\
q=1&\Blue{\bf1}\oplus\Blue{\bf3875}\oplus\Blue{\bf248}&{\bf 1}\oplus\Blue{\bf248}& {\bf248}\oplus\Blue{\bf1}
       &{\bf1}\oplus{\bf3875}\oplus{\bf248}\\
q=0&{\bf248}\oplus\Blue{\bf1}\oplus\Blue{\bf3875}\oplus\Blue{\bf248}
&{\bf 1}\oplus{\bf248}\oplus{\bf 1}
           & {\bf248}&{\bf1}\oplus{\bf3875}\\
q=-1 & {\bf248} & {\bf 1} 
  }
\end{align*}
  \caption{\it Part of the decomposition of the tensor hierarchy algebra
    $W(E_9)$ into $E_8$
  modules. The modules not present in ${\scr B}(E_9)$ are marked
  blue. Note the presence of $\tR_1={\bf1}$.}
\label{E8Table}
\end{table}

\section*{Acknowledgements}
We would like to thank David Berman and Charles Strickland-Constable
for collaboration in an early stage of this project. We also
acknowledge discussions with Alex Arvanitakis, Klaus Bering,
Olaf Hohm and Barton Zwiebach. This research is supported by the
Swedish Research Council, project no. 2015-04268. 

\bibliographystyle{utphysmod2}



\providecommand{\href}[2]{#2}\begingroup\raggedright\endgroup

\end{document}